**Exact Multivariate Two-Sample Density-Based Empirical Likelihood Ratio Tests Applicable to Retrospective and Group Sequential Studies**

Ablert Vexler[a], Gregory Gurevich[b] and Li Zou[c]

[a]*Department of Biostatistics, The State University of New York at Buffalo, Buffalo, NY 14214, U.S.A, avexler@buffalo.edu*

[b]*Department of Industrial Engineering and Management, SCE- Shamoon College of Engineering, Ashdod, Israel*

[c]*Department of Statistics and Biostatistics, California State University, East Bay, Hayward, CA 94542, U.S.A.*

ABSTRACT

Nonparametric tests for equality of multivariate distributions are frequently desired in research. It is commonly required that test-procedures based on relatively small samples of vectors accurately control the corresponding Type I Error (TIE) rates. Often, in the multivariate testing, extensions of null-distribution-free univariate methods, e.g., Kolmogorov-Smirnov and Cramér-von Mises type schemes, are not exact, since their null distributions depend on underlying data distributions. The present paper extends the density-based empirical likelihood technique in order to nonparametrically approximate the most powerful test for the multivariate two-sample (MTS) problem, yielding an exact finite-sample test statistic. We rigorously establish and apply one-to-one-mapping between the equality of vectors' distributions and the equality of distributions of relevant univariate linear projections. In this framework, we prove an algorithm that simplifies the use of projection pursuit, employing only a few of the infinitely many linear combinations of observed vectors' components. The displayed distribution-free strategy is employed in retrospective and group sequential manners. The asymptotic consistency of the proposed technique is shown. Monte Carlo studies demonstrate that the proposed procedures exhibit extremely high and stable power characteristics across a variety of settings. Supplementary materials for this article are available online.

## 1. Introduction

In many practical studies, data consist of multiple outcomes in the forms of random vectors realizations. We consider an illustrative example based on data from a study that has evaluated the association between biomarkers and myocardial infarction (MI). The study was focused on the residents of Erie and Niagara counties, 35-79 years of age (Schisterman et al., 2001). The New York State department of Motor Vehicles drivers' license rolls was used as the sampling frame for adults between the age of 35 and 65 years, while the elderly sample (age 65-79) was randomly chosen from the Health Care Financing Administration database. This research has examined the diagnostic ability of the biomarkers "thiobarbituric acid-reactive substances" (TBARS), "vitamin E", "glucose" and "high-density lipoprotein (HDL)-cholesterol", using samples that were collected on cases, who recently survived on MI disease, and controls, who had no previous MI disease. This practical issue requires formal developing of an MTS, "case-control", test in a retrospective manner. The difficulty in analyzing the TBARS, "vitamin E" and "glucose" biomarkers is that values of these biomarkers have been shown to be dependent and non-normal distributed.

A common approach in statistics is to generalize univariate testing mechanisms in multivariate setting. For example, the Hotelling $T^2$ test successfully extends Student's *t*-test to an MTS location decision-making strategy, when data follow a multivariate normal distribution. A departure of the underlying data distribution from being normal can imply a critical issue in controlling the TIE rate of the Hotelling $T^2$ test. The classical Hotelling $T^2$ test cannot be applied when the dimension of observed vectors exceeds the sample size (Biswas and Ghosh, 2014).

Perhaps, the difficulty to develop an exact test in the multivariate setting is due to the fact that, in general, the empirical estimator of the multivariate distribution function is not distribution free as in the univariate case (Simpson, 1951). Note also that Kolmogorov-Smirnov and Cramér-von Mises type two sample procedures may significantly suffer from a lack of power under various alternatives or when the size of one sample is relatively greater than the other sample (Gordon and Klebanov, 2010).

We can find within the modern statistical literature a line of research around constructions of MTS tests. For example, Biswas and Ghosh (2014) developed an MTS test based on inter-point distances that can be conveniently used in the high-dimensional low sample size setting (see also, e.g., Baringhaus and Franz, 2004, in this context). The authors showed the testing method that is asymptotically exact and consistent under general alternatives, when the sample size tends to infinity.

Jurečková and Kalina (2012) derived nonparametric MTS rank tests, considering location/scale alternatives. The special attention to distances associated with the samples' means and dispersion matrices afforded the authors to focus on Wilcoxon test type mechanisms (see also Marozzi,2016, in this context). Jurečková and Kalina (2012) concentrated on the MTS Wilcoxon, Psi and Savage exact (distribution-free) tests, taking into account some alternatives of Lehman type. The restriction that the dimension of the data is smaller than the sample size is employed in several rank tests' constructions presented in Jurečková and Kalina (2012: e.g., p.235).

Zhou et al. (2017) developed smooth MTS tests for special alternatives. In this development, the idea of projection pursues based on linear combinations of observed vectors' components was applied, reducing the MTS testing statement to the univariate problem. The authors assumed that observed samples contain realizations of independent *p*-dimensional vectors $\mathbf{X} = \left(X_1, ..., X_p\right)^T$ and $\mathbf{Y} = \left(Y_1, ..., Y_p\right)^T$, when every linear combination of $\mathbf{X}$'s components and every linear combination of $\mathbf{Y}$'s components distributed with respect to a broad family of parametric exponential type forms that can be formulated via $d$ parameters (see Sections 1.1 and 2.1 of Zhou et al., 2017, for details). Then allowing the number of parameters, $d$, to tend to infinity (along with the sample size), Zhou et al. (2017) proposed the test strategy for a large array of alternatives. The consideration of large sample sizes is essential in Zhou et al. (2017)'s developments. Assumption 4.3. of Zhou et al. (2017) restricts the sample sizes to be exceedingly superior to the vectors dimension, $p$. Unfortunately, the assumption that distributions of every linear combinations of vector components can be approximated by parametric distribution functions has not a direct interpretation in terms of restrictions related to the multivariate distribution of the vector, in general. We can also anticipate a difficulty in practical implementations of the proposed smooth tests when every linear combinations of vectors components should be employed in computations of the test statistics presented by Zhou et al. (2017) (see Shao and Zhou, 2010, in this context). In this case, evaluations based on all subsets $(r < p)$ of the vectors components $X_1,\ldots,X_p$ and $Y_1,\ldots,Y_p$ together with the use of an infinity number of linear combinations of them may not insure the correct decision regarding the equality of vectors $\mathbf{X}$ and $\mathbf{Y}$ distributions (Hamedani, 1984).

Zhou et al. (2017) proposed to substitute the null hypothesis $H_0$: "$\mathbf{X}$ and $\mathbf{Y}$ are identically distributed", say $\mathbf{X} \underset{d}{=} \mathbf{Y}$, by the hypothesis $\tilde{H}_0$: "$\mathbf{u}^T\mathbf{X} \underset{d}{=} \mathbf{u}^T\mathbf{Y}$, for all vectors $\mathbf{u}$ in the unit *p-1* dimensional sphere", applying the union-intersection principle. For a review, discussion and limitations related to the union-intersection principle, we can refer the reader to Olkin and Tomsky (1981). Perhaps, due to a possible strong dependence between sub-hypotheses contained in $\tilde{H}_0$ and their infinity number, the form of the null hypothesis applied in Zhou et al. (2017) needs a substantial analysis.

In the present paper, we provide rigorous probabilistic arguments to characterize the equality of vectors' distributions via the equality of distributions of relevant univariate linear projections. This implies constructing of precise multivariate extensions to powerful techniques developed in the univariate two-sample settings. Note that, developments and correct applications of univariate characterizations of vector distributions through the use of linear combinations of the vector components are not straightforward tasks even when observations are normally distributed (e.g., Hamedani, 1984; Shao and Zhou, 2010; Vexler, 2020).

Zhou et al. (2017) provided the multiplier bootstrap method to compute the critical value of the MTS smooth tests, which are not exact. According to Zhou et al. (2017), the limiting null distributions of the MTS smooth test statistics may not exist. We can remark that the smooth testing technique may suffer from a lack of power when underlying data are from nonexponential distributions (Vexler et al., 2014a).

In general, in the nonparametric MTS framework, there are no most powerful decision-making mechanisms. Although several techniques for the MTS problem have been proposed, there is still a demand for MTS tests developments based on strong statistical paradigms.

The main aim of this paper is to develop an exact and consistent approach for MTS testing of general alternatives, rigorously evaluating an applicability of the univariate projection pursuit in the MTS decision-making statement. The proposed testing strategy is derived without the assumption that the sample size exceeds the vectors dimension, *p.*

Note that, according to Friedman (1987), since linear projections are the simplest and most interpretable dimension-reducing methods, they are among most commonly used in theoretical and practical multivariate exploratory data analysis procedures. In practice, technical reasons restrict the number of the linear combinations related to projection pursuit to be considered. Then, it is important to show that a procedure supposedly based on the infinitely many linear combinations of multivariate variates can be conducted correctly by using a few relevant linear combinations of components. In the present paper we prove an MTS testing algorithm based on realizations of the linear combinations $\mathbf{u}^T\mathbf{X}$ and $\mathbf{u}^T\mathbf{Y}$ with selected finite values of the vector $\mathbf{u}$'s components.

In order to construct the proposed MTS test, we employ the density-based empirical likelihood ratio technique. The likelihood ratio methodology provides a basis for many important procedures and methods in statistical inference. By virtue of the Neyman-Pearson lemma, when functional forms of data distributions are completely specified the parametric likelihood approach is unarguably the most powerful tool. The parametric likelihood methods cannot be applied properly if assumptions on the forms of distributions of data do not hold. In this paper, we use a distribution-free strategy to approximate an optimal parametric likelihood ratio test-statistic via an empirical likelihood methodology.

Empirical likelihood (EL) concepts were introduced as nonparametric alternatives to parametric likelihood methods. The EL principle has been dealt with extensively across a variety of settings (e.g., Owen, 2001; Vexler *et al.*, 2014b, 2016). Commonly, the EL function has the form $EL = \prod_{i=1}^{n} p_i$, where the probability weights, $p_i,\ i = 1,...,n$ satisfy the assumptions $\sum_{i=1}^{n} p_i = 1$, $0 < p_i < 1,\ i = 1,\dots,n$ and the values of $p_i,\ i = 1,...,n$, are derived by maximizing the $EL$ function under empirical constraints. For example, when we draw a univariate sample of iid data points $Z_1,\dots,Z_n$ under the null hypothesis, $H_0$, that $\mathrm{E}(Z_1) = 0$, the corresponding constraint is $\sum_{i=1}^{n} p_i Z_i = 0$, an empirical version of the $H_0$-statement.

The density-based EL (DBEL) approach can represent nonparametric test statistics that approximate parametric Neyman-Pearson statistics (e.g., Nanda and Chowdhury, 2020; Vexler *et al.*, 2014a; Gurevich and Vexler, 2011). The DBEL method proposes to consider the likelihood in the form of $L_f = \prod_{i=1}^{n} f(Z_i) = \prod_{i=1}^{n} f_i,\ f_i = f\left(Z_{(i)}\right)$, where $f(\cdot)$ is a density function of observations $Z_1,...,Z_n$, and $Z_{(1)} \le \dots \le Z_{(n)}$ are the corresponding order statistics. The DBEL approach then approximates values of $f_j$ via maximization of $L_f$ given a constraint related to the empirical version of the density property of the form $\int f(u)du = 1$.

We introduce an MTS DBEL ratio decision-making mechanism with high and stable power properties for detecting general cases of nonequalities of two vectors' distributions. The proposed method is null-distribution-free, robust to model structures and highly efficient. This approach is applied in the retrospective setting with fixed sample sizes, as well as in the group sequential manner (e.g., Jennison and Turnbull, 1999; Zou, et al., 2019).

For the last 30 years, there has been growing interest in the use of group sequential designs in clinical studies. This is because such designs allow early stopping for either efficacy or futility, reducing the cost associated with data collections. In this context, for an extensive review and examples related to the group sequential methodology and its applications, we refer the reader to Jennison and Turnbull (1999). According to Jennison and Turnbull (1999), sequential comparisons of multiple outcomes' distributions in clinical trials belong to a first cohort of biostatistical targets. Thus, the MTS testing method developed in this paper is a valuable addition to statistical inference.

Commonly, sequential multivariate statistical procedures require to assume parametric forms of the underlying data distributions. Performances of parametric sequential tests strongly depend on the correctness of the distribution assumptions. Retrospective, non-sequential, studies are generally based on already collected datasets. In contrast to the analysis of data obtained retrospectively, we have the following issues related to sequential analysis. It can be difficult to specify the parametric distribution forms of the underlying data distributions before data points are observed. In a case we have strong reasons to assume the parametric forms of the data distributions, it could be extremely difficult, e.g., to test the corresponding parametric assumptions after the execution of sequential procedures. Sequential tests are based on random numbers of observations, and then data obtained after sequential analyses cannot be evaluated for goodness-of-fit using the conventional retrospective testing methods. In this paper we focus on a nonparametric MTS sequential procedure.

The paper is organized as follows. In Section 2, we characterize the multivariate statement of $\mathbf{X} \underset{d}{=} \mathbf{Y}$ by studying the distributions of the univariate projections of $\mathbf{X}$ and $\mathbf{Y}$. The DBEL method is introduced and extended to test for the MTS hypothesis. To simplify and correctly use the proposed decision-making mechanism, we prove an algorithm for computing extreme values of the MTS DBEL ratio test statistic that theoretically employs infinitely many linear combinations of underlying components of multivariate observations. This result can be associated with methods for conducting multivariate procedures developed via a Kolmogorov Smirnov type manner or ranks based concepts, when the univariate projection pursuit is employed. In Sections 3 and 4, the proposed method is used for constructing the retrospective and group sequential MTS tests. The asymptotic consistency of the developed tests is presented. We present exact mechanisms to control the TIE rates of the MTS DBEL ratio tests. An extensive Monte Carlo comparison between the proposed testing scheme and the modern MTS procedures is shown in Section 5. We discuss the performance of the tests under various alternative designs involving, e.g., cases when observed $\mathbf{X}$'s and $\mathbf{Y}$'s components are identically distributed, whereas $\mathbf{X}$ and $\mathbf{Y}$ have different distributions. It turns out that the proposed method significantly outperforms the known tests in almost all of the scenarios we considered. In Section 6, the real-world applicability of the proposed decision-make technique is illustrated using data from a study of biomarkers associated with myocardial infarction.

We conclude with remarks in Section 7. Proofs of the theoretical results and algorithms presented in this paper are outlined in the supplementary materials.

## 2. Method

This section displays the main stages in the MTS DBEL test statistic development. We begin by considering the statement that the observed data consists of two samples of realizations of independent *p*-dimensional random vectors $\mathbf{X} = \left(X_1, ..., X_p\right)^T$ and $\mathbf{Y} = \left(Y_1, ..., Y_p\right)^T$ with unknown multivariate distribution functions $F_X$ and $F_Y$, say $\mathbf{X} \sim F_X$ and $\mathbf{Y} \sim F_Y$, respectively. We test the null hypothesis $H_0 : F_X \equiv F_Y$ against the general alternative hypothesis $H_1 : F_X \neq F_Y$.

To reduce the dimension of the MTS testing problem, we consider the univariate linear combinations $X(\mathbf{u}) = \mathbf{u}^T \mathbf{X} = \sum_{i=1}^{p} u_i X_i$ and $Y(\mathbf{u}) = \mathbf{u}^T \mathbf{Y} = \sum_{i=1}^{p} u_i Y_i$, where vector $\mathbf{u} = \left(u_1,...,u_p\right)^T$.

Assuming that values of $\mathbf{X}$ and $\mathbf{Y}$ are measured on a continuous scale, we have the following characterization.

**Proposition 1.** The joint distribution functions $F_X$ and $F_Y$ are equal if and only if $X(\mathbf{u})$ and $Y(\mathbf{u})$ are identically distributed for all $u_1,...,u_p \in R^1$.

**Proof.** The proof treats the corresponding characteristic functions and is outlined in the supplementary materials.

To simplify the application of Proposition 1 to the further test development, we represent Proposition 1 in the form below.

**Proposition 2.** The next two statements are equivalent: (a) $F_X \equiv F_Y$, and (b) $X(\mathbf{u})$ and $Y(\mathbf{u})$ are identically distributed in each of the following scenarios regarding values of $u_j, j = \{1,...,, p\}$, selections: $A_s = \left\{\left(u_1,...,u_p\right): \text{ for } s > 1,\ u_1 = ... = u_{s-1} = 0;\ u_s = 1;\ u_j \in R^1, j = s+1,.., p\right\}$, $s = 1,..., p-1$, and $A_p = \left\{\left(u_1,...,u_p\right): u_i = 0, u_p = 1,\ i = 1,..., p-1\right\}$.

These results allow to use univariate outcomes from $X(\mathbf{u})$ and $Y(\mathbf{u})$ in constructing the test statistic for $H_0 : F_X \equiv F_Y$.

### 2.1 Development of the test statistic

We begin by outlining basic ingredients of the univariate two-sample DBEL test construction, introducing a principle of notations we use in this paper.

Assume we observe two independent samples $\{Z_1,...,Z_n\}$ and $\{V_1,...,V_m\}$, where $Z_1,...,Z_n \in R^1$ are iid data points, $V_1,...,V_m \in R^1$ are iid data points and $n, m$ are the sample sizes. Define the order statistics $Z_{(1)} \le ... \le Z_{(n)}$ and $V_{(1)} \le ... \le V_{(m)}$ based on the observations $\{Z_1,...,Z_n\}$ and $\{V_1,...,V_m\}$, respectively.

In order to test for the hypothesis $H_0 : F_Z \equiv F_V$, where $F_Z$ and $F_V$ denote the unknown distribution functions of $Z_1$ and $V_1$, respectively, Gurevich and Vexler (2011) developed the nonparametric DBEL approach for approximating the likelihood ratio $L = \prod_{i=1}^{n} f_Z\left(Z_{(i)}\right) \prod_{i=1}^{m} f_V\left(V_{(i)}\right) \left\{\prod_{i=1}^{n} f\left(Z_{(i)}\right) \prod_{i=1}^{m} f\left(V_{(i)}\right)\right\}^{-1}$, where density functions $f_Z$, $f_V$, $f$ correspond to $F_Z$, $F_V$, and an $H_0$-distribution of $Z_1$ and $V_1$, respectively. In the DBEL approach, values of $f_{Zi} = f_Z\left(Z_{(i)}\right), i \in \{1,...,n\}$, and $f_{Vj} = f_V\left(V_{(j)}\right), j \in \{1,...,m\}$, can be estimated by maximizing the likelihoods $\prod_{i=1}^{n} f_{Zi} \prod_{i=1}^{m} f_{Vi}$, given empirical constraints to control the assumptions $\int f(u) f_D(u) / f(u) du = 1, D \in \{Z,V\}$. According to Gurevich and Vexler (2011), for all integers $r < n/2$ and $s < m/2$, the corresponding empirical constrains have the forms

$$1 \ge \int_{Z_{(1)}}^{Z_{(n)}} \frac{f_Z(u)}{f(u)} f(u) du \cong \frac{1}{2r} \sum_{i=1}^{n} \frac{f_{Zi}}{f\left(Z_{(i)}\right)} \Delta_{Zir} \text{ and } 1 \ge \int_{V_{(1)}}^{V_{(m)}} \frac{f_V(u)}{f(u)} f(u) du \cong \frac{1}{2s} \sum_{i=1}^{m} \frac{f_{Vi}}{f\left(V_{(i)}\right)} \Delta_{Vis},$$

where $\Delta_{Zir} = \left\{F_{n+m}\left(Z_{(i+r)} \mid Z,V\right) - F_{n+m}\left(Z_{(i-r)} \mid Z,V\right)\right\}$,

$\Delta_{Vis} = \left\{F_{n+m}\left(V_{(i+s)} \mid Z,V\right) - F_{n+m}\left(V_{(i-s)} \mid Z,V\right)\right\}$; $Z_{(i+r)} = Z_{(n)}$, if $i+r > n$; $Z_{(i-r)} = Z_{(1)}$, if $i-r < 1$; $V_{(i+s)} = V_{(m)}$, if $i+s > m$; $V_{(i-s)} = V_{(1)}$, if $i-s < 1$;

$$F_{n+m}(t \mid Z,V) = (n+m)^{-1} \left\{\sum_{i=1}^{n} I\left(Z_i \le t\right) + \sum_{j=1}^{m} I\left(V_j \le t\right)\right\}$$

defines the $H_0$-empirical distribution function based on $Z_1,...,Z_n$ and $V_1,...,V_m$; and $I(.)$ is the indicator function. Then, the method of Lagrange multipliers yields the DBEL estimator of the likelihood ratio $L$ in the form $\prod_{i=1}^{n} 2r\left(n\Delta_{Zir}\right)^{-1} \prod_{j=1}^{m} 2s\left(m\Delta_{Vjs}\right)^{-1}$. The DBEL method incorporates an aspect of maximum likelihood methodology to state the test statistic

$$TS_{nm}(Z,V) = ELR_{Z,n} ELR_{V,m},$$

$$ELR_{Z,n} = \min_{a_n \le r \le b_n} \prod_{i=1}^{n} 2r\left(n\Delta_{Zir}\right)^{-1}, ELR_{V,m} = \min_{a_m \le s \le b_m} \prod_{i=1}^{m} 2s\left(m\Delta_{Vis}\right)^{-1},\ a_j = j^{0.5+\delta},$$

$$b_j = \min\left(j^{1-\delta}, 0.5j\right),\ j = n, m,\ \delta \in \left(0, 0.25\right).$$

Thus, by virtue of Propositions 1 and 2, in order to test for $H_0 : F_X \equiv F_Y$ based on observed vectors ${}_i\mathbf{X} = \left(X_{1i}, ..., X_{pi}\right)^T$, $i = 1, ..., n$, and ${}_j\mathbf{Y} = \left(Y_{1j}, ..., Y_{pj}\right)^T$, $j = 1, ..., m$, realizations of $\mathbf{X} = \left(X_1, ..., X_p\right)^T \sim F_X$ and $\mathbf{Y} = \left(Y_1, ..., Y_p\right)^T \sim F_Y$, we can focus on the statistic

$$TS_{nm}\left(X(\mathbf{u}), Y(\mathbf{u})\right) = ELR_{X(\mathbf{u}),n} ELR_{Y(\mathbf{u}),m} \quad \text{with}$$

$$ELR_{X(\mathbf{u}),n} = \min_{a_n \le r \le b_n} \prod_{i=1}^{n} 2r\left(n\Delta_{X(\mathbf{u})ir}\right)^{-1},\ ELR_{Y(\mathbf{u}),m} = \min_{a_m \le s \le b_m} \prod_{i=1}^{m} 2s\left(m\Delta_{Y(\mathbf{u})is}\right)^{-1},$$

where $\Delta_{X(\mathbf{u})ir} = \left\{F_{n+m}\left(X_{(i+r)}(\mathbf{u}) \mid X(\mathbf{u}), Y(\mathbf{u})\right) - F_{n+m}\left(X_{(i-r)}(\mathbf{u}) \mid X(\mathbf{u}), Y(\mathbf{u})\right)\right\}$,

$\Delta_{Y(\mathbf{u})is} = \left\{F_{n+m}\left(Y_{(i+s)}(\mathbf{u}) \mid X(\mathbf{u}), Y(\mathbf{u})\right) - F_{n+m}\left(Y_{(i-s)}(\mathbf{u}) \mid X(\mathbf{u}), Y(\mathbf{u})\right)\right\}$; $X_j(\mathbf{u}) = \sum_{i=1}^{p} u_i X_{ij}$, $j = 1, ..., n$, $Y_k(\mathbf{u}) = \sum_{i=1}^{p} u_i Y_{ik}$, $k = 1, ..., m$; $X_{(1)}(\mathbf{u}) \le ... \le X_{(n)}(\mathbf{u}), Y_{(1)}(\mathbf{u}) \le ... \le Y_{(m)}(\mathbf{u})$;

$F_{n+m}\left(t \mid X(\mathbf{u}), Y(\mathbf{u})\right) = \left(n+m\right)^{-1}\left\{\sum_{i=1}^{n} I\left(X_i(\mathbf{u}) \le t\right) + \sum_{j=1}^{m} I\left(Y_j(\mathbf{u}) \le t\right)\right\}$; $X_{(i+r)}(\mathbf{u}) = X_{(n)}(\mathbf{u})$, if $i + r > n$; $X_{(i-r)}(\mathbf{u}) = X_{(1)}(\mathbf{u})$, if $i - r < 1$; $Y_{(i+s)}(\mathbf{u}) = Y_{(m)}(\mathbf{u})$, if $i + s > m$; $Y_{(i-s)}(\mathbf{u}) = Y_{(1)}(\mathbf{u})$, if $i - s < 1$. Note that, defining the function $g_k\left(t\right) = I\left(t \le 0\right) + tI\left(0 < t \le k\right) + kI(t > k)$, we can simplify the notations

$$F_{n+m}\left(X_{(i)}(\mathbf{u}) \mid X(\mathbf{u}), Y(\mathbf{u})\right) = \left(n+m\right)^{-1}\left\{g_n\left(i\right) + \sum_{j=1}^{m} I\left(Y_j(\mathbf{u}) \le X_{(i)}(\mathbf{u})\right)\right\},$$

$$F_{n+m}\left(Y_{(j)}(\mathbf{u}) \mid X(\mathbf{u}), Y(\mathbf{u})\right) = \left(n+m\right)^{-1}\left\{\sum_{i=1}^{n} I\left(X_i(\mathbf{u}) \le Y_{(j)}(\mathbf{u})\right) + g_m\left(j\right)\right\}.$$

In order to make decision regarding the MTS problem, it is reasonable to employ a concept associated with the Kolmogorov-Smirnov principle for measuring distances between nonparametric hypotheses. To this end, we can use large values of

$$TS_{nm} = \max_{(u_1,...,u_p)\in\bigcup_{j=1}^p A_j} TS_{nm}\left(X(\mathbf{u}),Y(\mathbf{u})\right),$$

discriminating $H_0$ and its alternative hypothesis.

The requirement for computing the maximum of the statistic $TS_{nm}\left(X(\mathbf{u}),Y(\mathbf{u})\right)$ over all $u_1,...,u_p \in R^1$ is not at least user-friendly. Note, for example, that, commonly, to implement Kolmogorov-Smirnov type statistics, algorithms for conducting maximums involved in the statistics can be performed by using finite numbers of arguments based on observations. Towards this, the following results are obtained.

For clarity of explanation, we begin by exemplifying the case with $p = 2$, where we are interested in conducting the statistic $TS_{nm} = \max_{(u_1,u_2)\in A_1\bigcup A_2} TS_{nm}\left(X(\mathbf{u}),Y(\mathbf{u})\right)$. The sets $A_1 = \left\{(u_1,u_2):\ u_1 = 1, u_2 \in R^1\ \right\}$ and $A_2 = \left\{(u_1,u_2):\ u_1 = 0, u_2 = 1\ \right\}$ are defined in Proposition 2. Define the data points $W(i,j) = \left(X_{1i} - Y_{1j}\right)\left(Y_{2j} - X_{2i}\right)^{-1}$, $U(i,r) = \left(X_{1i} - X_{1r}\right)\left(X_{2r} - X_{2i}\right)^{-1}$ and $V(j,s) = \left(Y_{1j} - Y_{1s}\right)\left(Y_{2s} - Y_{2j}\right)^{-1}$ in order to show that

$$TS_{nm} = \max_{(u_1,u_2)\in B_1\bigcup A_2} TS_{nm}\left(X(\mathbf{u}),Y(\mathbf{u})\right),$$

where the set

$B_1 = \left[(u_1,u_2): u_1 = 1, u_2 \in \left\{W(i,j), U(i,r), V(j,s),\ i \neq r, j \neq s, 1 \le i,r \le n,\ 1 \le j,s \le m\right\}\right]$.

To establish this result, we will apply the following scheme. The statistic $TS_{nm}\left(X(\mathbf{u}),Y(\mathbf{u})\right)$ is based on the variables $I\left\{X_i(\mathbf{u}) \le Y_{(j)}(\mathbf{u})\right\}$, $I\left\{Y_j(\mathbf{u}) \le X_{(i)}(\mathbf{u})\right\}$, $I\left\{X_i(\mathbf{u}) \le X_{(k)}(\mathbf{u})\right\}$, $I\left\{Y_j(\mathbf{u}) \le Y_{(r)}(\mathbf{u})\right\}$, $1 \le i,k \le n,\ 1 \le j,r \le m$. Thus, two vectors $\mathbf{u} = (u_1,u_2)^T$, $\mathbf{v} = (v_1,v_2)^T$ satisfy $TS_{nm}\left(X(\mathbf{u}),Y(\mathbf{u})\right) = TS_{nm}\left(X(\mathbf{v}),Y(\mathbf{v})\right)$, if $I\left\{X_i(\mathbf{u}) \le Y_j(\mathbf{u})\right\} = I\left\{X_i(\mathbf{v}) \le Y_j(\mathbf{v})\right\}$, $I\left\{X_i(\mathbf{u}) \le X_k(\mathbf{u})\right\} = I\left\{X_i(\mathbf{v}) \le X_k(\mathbf{v})\right\}$, $I\left\{Y_j(\mathbf{u}) \le Y_r(\mathbf{u})\right\} = I\left\{Y_j(\mathbf{v}) \le Y_r(\mathbf{v})\right\}$, for all $1 \le i,k \le n$ and $1 \le j,r \le m$. In the supplementary materials, we display details of the proof that, for $\mathbf{u} \in A_1 \cup A_2$ and $\mathbf{v} \in B_1 \cup A_2$, $I\left\{X_i(\mathbf{u}) \le Y_j(\mathbf{u})\right\} = I\left\{X_i(\mathbf{v}) \le Y_j(\mathbf{v})\right\}$,

$I\left\{X_i(\mathbf{u}) \le X_k(\mathbf{u})\right\} = I\left\{X_i(\mathbf{v}) \le X_k(\mathbf{v})\right\}$, $I\left\{Y_j(\mathbf{u}) \le Y_r(\mathbf{u})\right\} = I\left\{Y_j(\mathbf{v}) \le Y_r(\mathbf{v})\right\}$, for all $1 \le i,k \le n$ and $1 \le j,r \le m$.

In order to consider the general case with dimension $p$, we recursively define the following notations. Let $J_{W,j} = \left(i_1, j_1, ..., i_{2^{j-1}}, j_{2^{j-1}}\right)$, $J^c_{W,j} = \left(i_{2^{j-1}+1}, j_{2^{j-1}+1}, ..., i_{2^j}, j_{2^j}\right)$, $J_{U,j} = \left(i_1, r_1, ..., i_{2^{j-1}}, r_{2^{j-1}}\right)$, $J^c_{U,j} = \left(i_{2^{j-1}+1}, r_{2^{j-1}+1}, ..., i_{2^j}, r_{2^j}\right)$, $J_{V,j} = \left(j_1, s_1, ..., j_{2^{j-1}}, s_{2^{j-1}}\right)$ and $J^c_{V,j} = \left(j_{2^{j-1}+1}, s_{2^{j-1}+1}, ..., j_{2^j}, s_{2^j}\right)$ denote integer row-vectors with the components $1 \le i_q \ne r_q \le n,\ 1 \le j_q \ne s_q \le m$, where $q = 1,...,2^j$ and $j = 1,...,p-1$. For example, $J_{W,3} = \left(i_1, j_1, i_2, j_2, i_3, j_3, i_4, j_4\right)$, $J^c_{W,3} = \left(i_5, j_5, ..., i_8, j_8\right)$. We can write $J_{W,1} = (i,j)$, $J_{U,1} = (i,r)$ and $J_{V,1} = (j,s)$, for the sake of simplicity. When $J_{W,j} \ne J^c_{W,j}$, $J_{U,j} \ne J^c_{U,j}$, $J_{V,j} \ne J^c_{V,j}$, we denote, for $k = 1,...,p-1$, the sets of the random variables

$W_k\left(J_{W,1}\right) = \left(X_{ki} - Y_{kj}\right)\left(Y_{pj} - X_{pi}\right)^{-1}$, $U_k\left(J_{U,1}\right) = \left(X_{ki} - X_{kr}\right)\left(X_{pr} - X_{pi}\right)^{-1}$,

$V_k\left(J_{V,1}\right) = \left(Y_{kj} - Y_{ks}\right)\left(Y_{ps} - Y_{pj}\right)^{-1}$, and then, for $j = 2,...,p-1$, $k = 1,...,p-j$,

$$W_k\left(J_{W,j}\right) = \left\{W_k\left(J_{W,j-1}\right) - W_k\left(J^c_{W,j-1}\right)\right\}\left\{W_{p-j+1}\left(J^c_{W,j-1}\right) - W_{p-j+1}\left(J_{W,j-1}\right)\right\}^{-1},$$

$$U_k\left(J_{U,j}\right) = \left\{U_k\left(J_{U,j-1}\right) - U_k\left(J^c_{U,j-1}\right)\right\}\left\{U_{p-j+1}\left(J^c_{U,j-1}\right) - U_{p-j+1}\left(J_{U,j-1}\right)\right\}^{-1},$$

$$V_k\left(J_{V,j}\right) = \left\{V_k\left(J_{V,j-1}\right) - V_k\left(J^c_{V,j-1}\right)\right\}\left\{V_{p-j+1}\left(J^c_{V,j-1}\right) - V_{p-j+1}\left(J_{V,j-1}\right)\right\}^{-1}.$$

Note, for example, that the notation $W_k\left(J_{W,1}\right)$ means the sequence $\left(X_{k1} - Y_{k1}\right)\left(Y_{p1} - X_{p1}\right)^{-1}$, $\left(X_{k1} - Y_{k2}\right)\left(Y_{p2} - X_{p1}\right)^{-1}$,..., $\left(X_{kn} - Y_{km}\right)\left(Y_{pm} - X_{pn}\right)^{-1}$.

**Proposition 3.** We have

$$TS_{nm} = \max_{\left(u_1,...,u_p\right)\in\bigcup_{j=1}^p A_j} TS_{nm}\left(X(\mathbf{u}), Y(\mathbf{u})\right) = \max_{\left(u_1,...,u_p\right)\in\bigcup_{j=1}^p B_j} TS_{nm}\left(X(\mathbf{u}), Y(\mathbf{u})\right),$$

where sets $B_1,...,B_p$ contain elements defined via the following algorithms: for $s = 1,...,p-1$,

$B_s = \left[ \left( u_1, ..., u_p \right) : \text{ for } s > 1,\ u_1 = ... = u_{s-1} = 0;\ u_s = 1;\ \text{for } d = 1, ..., p-s, \text{ given } u_1, ..., u_{s+d-1}, \text{ select} \right.$

$$\left. u_{s+d} \in \left\{ \sum_{h=s}^{s+d-1} W_h \left( J_{W,p-d-s+1} \right) u_h,\ \sum_{h=s}^{s+d-1} U_h \left( J_{U,p-d-s+1} \right) u_h,\ \sum_{h=s}^{s+d-1} V_h \left( J_{V,p-d-s+1} \right) u_h \right\} \right] \text{ and } B_p = A_p .$$

We exemplify details of the notations used in Proposition 3, when $p = 3$, in the supplementary materials.

The proof scheme for deriving the statement of Proposition 3 associates the indicators $I\left\{X_i(\mathbf{u}) \le Y_j(\mathbf{u})\right\}$, $I\left\{X_i(\mathbf{u}) \le X_j(\mathbf{u})\right\}$, $I\left\{Y_i(\mathbf{u}) \le Y_j(\mathbf{u})\right\}$ with $I\left\{X_i(\mathbf{v}) \le Y_j(\mathbf{v})\right\}$, $I\left\{X_i(\mathbf{v}) \le X_j(\mathbf{v})\right\}$, $I\left\{Y_i(\mathbf{v}) \le Y_j(\mathbf{v})\right\}$, for all $i, j$, $\mathbf{u} = \left(u_1, ..., u_p\right)^T \in \bigcup_{j=1}^{p} A_j$ and $\mathbf{v} = \left(v_1, ..., v_p\right)^T \in \bigcup_{j=1}^{p} B_j$. This proof algorithm can be applied to execute different multivariate procedures developed by using a Kolmogorov Smirnov type manner or ranks based concepts, when the univariate projection pursuit is employed (see the supplementary materials, for details).

**Remark 1.** Various Monte Carlo experiments based on realizations of $\mathbf{X}, \mathbf{Y}$ with a variety of sample sizes *(n,m)* showed that multiple uses of the R function '*optim*' executed with initial values equating to different empirical quantiles of $\left(u_1, ..., u_p\right) \in B_k$, $k = 1, ..., p,$ can significantly reduce the computation time of the proposed procedure.

## 3. The MTS DBEL test based on pre-collected samples with nonrandom sizes

Often, practical studies consider retrospectively collected *p*-dimensional independent outcomes from two groups, say $\mathbf{X} = \left(X_1, ..., X_p\right)^T$ and $\mathbf{Y} = \left(Y_1, ..., Y_p\right)^T$, following multivariate distributions $F_X$ and $F_Y$, respectively. Let ${}_i\mathbf{X} = \left(X_{1i}, ..., X_{pi}\right)^T$ be the outcomes of the *i*-th subject. $i = 1, ..., n,$ from the $\mathbf{X}$-sample as well as let ${}_j\mathbf{Y} = \left(Y_{1j}, ..., Y_{pj}\right)^T$ be the outcomes of the *j*-th subject, $j = 1, ..., m,$ from the $\mathbf{Y}$-sample. The null hypothesis being tested is $H_0 : F_X \equiv F_Y$. Towards this end, we define the linear combinations $X_j(\mathbf{u}) = \mathbf{u}^T {}_j\mathbf{X} = \sum_{i=1}^{p} u_i X_{ij}, j = 1, ..n,$ and $Y_k(\mathbf{u}) = \mathbf{u}^T {}_k\mathbf{Y} = \sum_{i=1}^{p} u_i Y_{ik}$, obtaining the test statistic $TS_{nm}$. We then propose to reject $H_0$ if $TS_{nm} > C_\alpha$, where $C_\alpha$ is an $\alpha$-level test threshold.

It is clear that, the proposed test is exact. In Section 3.1 we discuss computing of the critical values of the proposed testing strategy.

The next result points out that the MTS DBEL procedure is an asymptotic power one test.

Let $\Pr_{H_k}$ and $\mathrm{E}_{H_k}$ denote the probability measure and expectation under $H_k, k = 0,1$. Denote the density functions $f_{H_0}(t;\mathbf{u}) = d\Pr_{H_0}\{X_1(\mathbf{u}) \le t\}/dt = d\Pr_{H_0}\{Y_1(\mathbf{u}) \le t\}/dt$ and

$f_{X(\mathbf{u})}(t;\mathbf{u}) = d\Pr_{H_1}\{X_1(\mathbf{u}) \le t\}/dt,\ f_{Y(\mathbf{u})}(t;\mathbf{u}) = d\Pr_{H_1}\{Y_1(\mathbf{u}) \le t\}/dt$ and assume that $n/m \to \eta$ as $n,m \to \infty$, where a constant $\eta > 0$. Then the following proposition shows the consistency of the proposed test.

**Proposition 4.** Let $F_X$ and $F_Y$ be absolute continuous distribution function defined on $R^p$.

Assume there exists a vector $\mathbf{u}_0 = \left(u_{01},...,u_{0p}\right)^T$ such that, for $\xi_0 = X_1(\mathbf{u}_0)$ and $\xi_0 = Y_1(\mathbf{u}_0)$, the expectations $\mathrm{E}\left[\log\left\{f_{H_0}(\xi_0;\mathbf{u}_0)\right\}\right]$, $\mathrm{E}\left[\log\left\{f_{X(\mathbf{u}_0)}(\xi_0;\mathbf{u}_0)\right\}\right]$ and $\mathrm{E}\left[\log\left\{f_{Y(\mathbf{u}_0)}(\xi_0;\mathbf{u}_0)\right\}\right]$ are finite. Then, for a positive threshold $C$, we have

$\Pr_{H_1}\left\{(n+m)^{-1}\log(TS_{nm}) > C\right\} \to 1$, whereas $\Pr_{H_0}\left\{(n+m)^{-1}\log(TS_{nm}) > C\right\} \to 0$, as $n,m \to \infty$.

**Proof.** The proof uses the theorem of Dvoretzky, Kiefer and Wolfowitz (Serfling, 2009, p. 59) and is outlined in the supplementary materials.

### 3.1. Null distribution

In this section, we point out that the proposed test statistic is exact, i.e., $H_0$-distribution-free. We then present the critical values for the proposed test for different sample sizes and *p=2* as well as an R code to derive the test critical values in practice.

Note that, under $H_0$, the test statistic $TS_{nm}$ depends only on the empirical distribution function $F_{n+m}\left(t \mid X(\mathbf{u}), Y(\mathbf{u})\right)$, for all $\mathbf{u} \in R^p$, which in turn depend only on certain indicator functions. By virtue of Proposition 1, under $H_0$, we have the distribution function $F(t) \equiv \Pr_{H_0}\{X_i(\mathbf{u}) \le t\} = \Pr_{H_0}\{Y_j(\mathbf{u}) \le t\}$, for all $\mathbf{u} \in R^p$ and $i \in \{1,...,n\}, j \in \{1,...,m\}$, Then, $I(X(\mathbf{u}) \le Y(\mathbf{u})) = I[F\{X(\mathbf{u})\} \le F\{Y(\mathbf{u})\}] = I(U_1 \le U_2)$, where $U_1$, $U_2$ are *Uniform*(0, 1) distributed. (For details regarding distribution-free test constructions based on empirical distribution functions see Crouse, 1966). Hence, it follows that

$$\Pr_{H_0}\{\log(TS_{nm}) > C_\alpha\}$$
$$= \Pr\left\{\log(TS_{nm}) > C_\alpha \mid {}_i\mathbf{X} = (X_{1i},...,X_{pi})^T, {}_j\mathbf{Y} = (Y_{1j},...,Y_{pj})^T \sim N(0, I_p), i = 1,...,n, j = 1,...,m\right\} = \alpha,$$

where $I_p$ is a $p$-dimensional identity matrix. Therefore, the proposed method is exact, and then the critical values for the proposed DBEL test can be accurately approximated using Monte Carlo techniques.

Note that extensive Monte Carlo simulations confirmed the robustness of the proposed test with respect to the values of $\delta \in (0, 0.25)$ at the definition of the test $\log(TS_{nm}) > C_\alpha$. For practical purposes, we suggest a value of $\delta = 0.1$ that is used in our applications. It was shown in the DBEL literature that power values of DBEL type test statistics do not differ substantially for values of $\delta \in (0, 0.25)$ (e.g., Gurevich and Vexler, 2011; Tsai et al., 2013; Vexler et al., 2014a).

In the case with *p=2*, we tabulated the percentiles of the null distribution of the test statistic based on 20,000 samples of ${}_i\mathbf{X} = (X_{1i}, X_{2i})^T \sim N(0, I_2)$, $i = 1,..,n$, and ${}_j\mathbf{Y} = (Y_{1j}, Y_{2j})^T \sim N(0, I_2)$, $j = 1,...,m$, to calculate values of $TS_{nm}$ at each pair (*n, m*). The generated values of the test statistic $\log(TS_{nm})$ were used to determine the critical values $C_\alpha$ of the null distribution of $\log(TS_{nm})$ at the significance level $\alpha$. The results of this Monte Carlo study are presented in Table 1.

**Table 1.** *Critical Values of the Proposed Test Statistic,* $\log(TS_{nm})$

| Significance level $\alpha$ | Significance level $\alpha$ |
|---|---|

| $(n,m)$ | **0.1** | **0.05** | **0.01** | $(n,m)$ | **0.1** | **0.05** | **0.01** |
|---|---|---|---|---|---|---|---|
| (10, 10) | 11.063 | 11.983 | 13.832 | (15, 40) | 15.592 | 16.538 | 18.397 |
| (10, 15) | 11.878 | 12.656 | 14.470 | (15, 50) | 16.400 | 17.304 | 19.311 |
| (10, 30) | 14.276 | 15.111 | 16.874 | (20, 20) | 13.877 | 14.714 | 16.560 |
| (10, 40) | 15.128 | 16.026 | 17.822 | (20, 30) | 15.226 | 16.108 | 18.029 |
| (10, 50) | 15.918 | 16.863 | 18.904 | (20, 40) | 16.112 | 17.021 | 18.978 |
| (15, 15) | 12.510 | 13.374 | 15.168 | (20, 50) | 16.896 | 17.853 | 19.855 |
| (15, 20) | 13.249 | 14.090 | 15.959 | (50,50) | 19.538 | 20.317 | 22.340 |
| (15, 30) | 14.714 | 15.612 | 17.475 | (65, 70) | 22.677 | 23.428 | 24.850 |

An R function (R Development Core Team, 2012) for the Monte Carlo computations related to the critical values $C_\alpha$ of the null distribution of $\log(TS_{nm})$ is shown in the supplementary materials. This function can be easily modified to prepare the proposed test in a real study.

We can remark that, since the statistic $\log(TS_{nm})$ is $H_0$-distribution-free, Monte Carlo or Bootstrap type procedures can be employed to estimate different probabilistic characteristics of $\log(TS_{nm})$, under $H_0$, e.g., $\mathrm{var}\{\log(TS_{nm})\}$.

## 4. The group sequential MTS DBEL test

Investigators can design a clinical study in group sequential manners that offer the possibility to stop an experiment early with a statistically significant test result. The sequential trial can require need less observations than the trial with fixed sample sizes in which a test decision can be made only at the end of the trial.

Following Jennison and Turnbull (2000), we assume that *K* groups of subjects are available from two sets, say set *A* and set *B*. Let ${}_i\mathbf{X} = (X_{1i},...,X_{pi})^T \sim F_X$ and ${}_i\mathbf{Y} = (Y_{1i},...,Y_{pi})^T \sim F_Y$, $i \ge 1$, be independent and represent the measures of subjects allocated to sets *A* and *B*, respectively. For simplicity, according to a constrained randomization scheme of the basic two-sample comparison, we assume *m* subjects from *A* and *m* subjects from *B* can provide their measurements in every group $k = 1,...,K$ . Consider the problem of testing the null hypothesis $H_0 : F_X \equiv F_Y$ , when the distribution functions $F_X, F_Y$ are unknown. For example, sets *A* and *B* can be used to indicate individuals who receive two different treatments. In this case, $H_0$ states that no treatment difference. To arrange the sequential decision-making procedure, say $SP_K$ , we fix a threshold $C_{\alpha,K}$ and execute the next algorithm: starting from $k = 1$ , (a) for group $k$ , compute the statistic $R_{km} = \max_{u_1,...,u_p} ELR_{X(\mathbf{u}),km} ELR_{Y(\mathbf{u}),km}$ ; (b) if $\log(R_{km}) \geq C_{\alpha,K}$ , stop and reject $H_0$ otherwise, if $k+1 \leq K$ , continue to step (a), employing group $k+1$ ; (c) in the case with $\log(R_{km}) < C_{\alpha,K}$ , for all $k = 1,...,K$ , do not reject $H_0$ . The critical value $C_{\alpha,K}$ can be chosen to give overall TIE $\alpha$ , i.e., $\Pr_{H_0}(SP_K \text{ rejects } H_0) = \alpha$ . In a similar manner to Section 3.1, we note that the null distribution of $\log(R_{km})$ does not depend on underlying data distributions. Then, for different values of $\alpha$ , e.g., $\alpha = 0.01, 0.05$, and $m$ , $C_{\alpha,K}$ can be computed numerically that is exemplified in Section 4.1.

Note that $\Pr(SP_K \text{ rejects } H_0) = \Pr\left[\bigcup_{j=1}^{K} \left\{\log(R_{jm}) \geq C_{\alpha,K}\right\}\right]$.

Proposition 5 below displays that the nonparametric procedure $SP_K$ is consistent.

**Proposition 5.** Let the assumptions of Proposition 4 be held and, for a constant $d > 0$ , $1 \leq K = O(m^d)$ as $m \to \infty$. Then, for a fixed $C > 0$ , we have

$\Pr_{H_1}\left\{\max_{1\leq k\leq K} \log(R_{km}) > mC\right\} \to 1$, whereas $\Pr_{H_0}\left\{\max_{1\leq k\leq K} \log(R_{km}) > mC\right\} \to 0$, as $m \to \infty$.

**Proof.** The proof uses the proof scheme of Proposition 4 and is shown in the supplementary materials.

**4.1. Null distribution**

In this section, we exemplify evaluations of the critical values $C_{\alpha,K}$ for the proposed procedure $SP_K$, considering different values of *K* and *m*. An R code for the corresponding computations is presented in the supplementary materials.

Since $\Pr_{H_0}\left\{\max_{1\le k\le K}\log\left(R_{km}\right)<C_{\alpha,K}\right\}=\Pr_{H_0}\left\{\log\left(R_{1m}\right)<C_{\alpha,K},...,\log\left(R_{Km}\right)<C_{\alpha,K}\right\}$, in a similar manner to Section 3.1, we can write the TIE rate of $SP_K$ in the form

$$\Pr_{H_0}\left\{SP_K \text{ rejects } H_0\right\}=\Pr\left\{\max_{1\le k\le K}\log\left(R_{km}\right)>C_{\alpha,K}\mid {}_i\mathbf{X},\, {}_j\mathbf{Y}\sim N(0,I_p), 1\le i,j\le Km\right\}=\alpha.$$

In the case with *p=2*, we tabulated the percentiles of the null distribution of the test statistic $\max_{1\le k\le K}\log\left(R_{km}\right)$ based on 20,000 generated samples of ${}_i\mathbf{X}=\left(X_{1i},X_{2i}\right)^T\sim N\left(0,I_2\right)$, and ${}_j\mathbf{Y}=\left(Y_{1j},Y_{2j}\right)^T\sim N(0,I_2)$, $1\le i,j\le Km$, when the number of groups *K*=2, 3, 4, 5, and the sample size per group *m*=5, 10. The results of this Monte Carlo study are displayed in Table 2.

**Table 2.** Critical values $C_{\alpha,K}$ of the group sequential MTS DBEL test based on *K* (=2, 3, 4, 5) groups of observations with *m* (=5, 10) data points per group. The significance level is $\alpha$ .

| *K* | *M* | $\alpha$=0.2 | $\alpha$=0.1 | $\alpha$=0.05 | $\alpha$=0.01 |
|---|---|---|---|---|---|
| 2 | 5 | 10.413 | 11.053 | 11.987 | 13.872 |
| | 10 | 13.067 | 13.918 | 14.737 | 16.452 |
| 3 | 5 | 11.818 | 12.660 | 13.525 | 15.410 |
| | 10 | 15.663 | 16.502 | 17.321 | 19.207 |
| 4 | 5 | 13.185 | 14.103 | 14.938 | 16.859 |
| | 10 | 17.277 | 18.136 | 19.004 | 20.864 |
| 5 | 5 | 14.458 | 15.338 | 15.338 | 18.018 |
| | 10 | 18.734 | 19.569 | 20.405 | 22.401 |

We can remark that, in the context of group sequential testing, Monte Carlo approximations to the critical values are often applied in practice (e.g., Jennison and Turnbull, 1999).

## 5. Monte Carlo Experiments

We carried out an extensive Monte Carlo study to explore the performance of the proposed testing strategy in comparison to the modern MTS tests developed by Jurečková and Kalina (2012), Zhou et al. (2017), Biswas and Ghosh (2014). We are grateful for the authors of the papers, Jurečková and Kalina (2012) and Zhou et al. (2017), who provided us relevant programming codes and/or discussed details regarding implementations of their tests. As suggested, we used the Jurečková and Kalina (2012) type MTS test in the form proposed by Marozzi (2016).

We considered generating data from various scenarios, evaluating more than 210 designs of sampling $\mathbf{X}$ and $\mathbf{Y}$, under $H_1$. In this study, a case, where the proposed method demonstrated a relatively weak power, was not detected. Table 3 displays Designs $D_1$,…, $D_9$ and their explanations, exemplifying the treated scenarios.

**Table 3.** *Distributions for* $\mathbf{X} = \left(X_1, X_2\right)^T$ *and* $\mathbf{Y} = \left(Y_1, Y_2\right)^T$ *used in the power study.*

| Designs | *Models/Descriptions* |
|---|---|
| $D_1$ | $X_1 \sim N\left(0,1\right), X_2 \sim N\left(0,1\right)$ vs. $\mathbf{Y}$ from a multivariate *t*-distribution with scale matrix $\begin{bmatrix} 1 & 0.9 \\ 0.9 & 1 \end{bmatrix}$ and degrees of freedom df=7, the corresponding R function (R Development Core Team, 2012) is 'rmvt'. |
| $D_2$ | $X_1 \sim N\left(0,1\right), X_2 \sim Exp(1)$ vs. $Y_1 \sim N\left(0,1\right), Y_2 \sim LogN\left(0,1\right)$, where $X_1, X_2, Y_1, Y_2$ are independent. $D_2$ uses the skewed distributions. |
| $D_3$ | $X_1 \sim Unif(-1,1), X_2 \sim N\left(0,1.5^2\right)$ vs. $Y_1 \sim N\left(0,1\right), Y_2 \sim N\left(0,1\right)$, where $X_1, X_2, Y_1, Y_2$ are independent. $D_3$ represents a case of a uniform distribution vs. a normal distribution as well as different variances of $\mathbf{X}$- and $\mathbf{Y}$-samples. (Oftentimes, it is |

difficult to goodness-of-fit-detect departures from normal distributions based on uniformly distributed data.)

D4 $X_1 \sim N(0,1), X_2 \sim N(0,1)$ vs. $Y_1 \sim N(0.5,1), Y_2 \sim N(0.7,1)$, where $X_1, X_2, Y_1, Y_2$ are independent. D4 shows an obvious location-difference in $\mathbf{X}$- and $\mathbf{Y}$-distributions.

D5 $\mathbf{X} \sim N(\boldsymbol{\mu}, \Sigma_X)$ vs. $\mathbf{Y} \sim N(\boldsymbol{\mu}, \Sigma_Y)$, where $\boldsymbol{\mu} = (0,0)^T, \Sigma_X = \begin{bmatrix} 1 & 0 \\ 0 & 1 \end{bmatrix}, \Sigma_Y = \begin{bmatrix} 1 & 0.5 \\ 0.5 & 1 \end{bmatrix}$.

D5 shows an obvious scale-difference in $\mathbf{X}$- and $\mathbf{Y}$-distributions.

D6 $\mathbf{X} \sim N(\boldsymbol{\mu}, \Sigma_X)$ vs. $\mathbf{Y} \sim N((0.5, 0.7)^T, \Sigma_Y)$, where $\boldsymbol{\mu}, \Sigma_X, \Sigma_Y$ are defined in D5. D6 shows a location/scale-difference in $\mathbf{X}$- and $\mathbf{Y}$-distributions.

D7 $\mathbf{X} \sim N(\boldsymbol{\mu}, \Sigma_X)$ vs. $Y_1 = \xi - 1, \xi \sim Exp(1), Y_2 \sim N(0,1)$, where $Y_1, Y_2$ are independent.

D8 $\mathbf{X} \sim N(\boldsymbol{\mu}, \Sigma_X)$ vs. $Y_1 \sim N(0,1)$, $Y_2 = \tau Y_1$, where random variable $\tau = -1$ or $1$ and is independent of $Y_1$, $Pr(\tau_1 = -1)$=0.5, $\boldsymbol{\mu}$ is defined above. In this case, $Y_2 \sim N(0,1)$, but $\mathbf{Y}$ is not bivariate normal.

D9 $\mathbf{X} \sim N(\boldsymbol{\mu}, \Sigma_X)$ vs. $Y_1 = \xi^{1/2}\eta_1 + (1-\xi)^{1/2}\eta_2, Y_2 = \xi^{1/2}\eta_3 + (1-\xi)^{1/2}\eta_2$, where $\xi \sim Unif(0,1)$ and $\eta_1, \eta_2, \eta_3$ are independent $N(0,1)$-distributed random variables, $\boldsymbol{\mu}$ is defined in D5. In this case, $Y_1 \sim N(0,1), Y_2 \sim N(0,1)$, but $\mathbf{Y}$ is not bivariate normal, Stoyanov (2014), p. 97-98.

---

Assuming that reasonable MTS tests based on large samples provide relatively equivalent and powerful outputs, in this Monte Carlo study, we focused on 10,000 replicates of $\{{}_1\mathbf{X}, ..., {}_n\mathbf{X}\}$ and $\{{}_1\mathbf{Y}, ..., {}_m\mathbf{Y}\}$ for each pair $(n,m) =$ (10,15), (15,10), (30,50), (50,30), (50,50), (65,70) and (70,65).

Table 4 shows the results of the power evaluations of the proposed test (“$Log(TS_{nm})$”), Zhou et al. (2017)’s test (“Z”), Jurečková and Kalina (2012)’s and Marozzi (2016)’s test (“JKM”) as well as Biswas and Ghosh (2014)’s test (“BG”). The significance level of the tests, $\alpha$, was supposed to be fixed at 5%. This study demonstrates the DBEL MTS test is superior to the considered Z, JKM and BG tests in almost all scenarios under the designs of $D_1,\dots, D_9$. Only in the case $D_2$ with $(n,m)$=(10,15), the Monte Carlo estimated power of $Log(TS_{nm})$ is less than that that of BG as 0.1049<0.1057, respectively. In contrast, under $D_2$ with $(n,m)$=(30,50), the DBEL test and BG test have power levels of 0.3205 and 0.1265, respectively.

**Table 4.** *The Monte Carlo power of the tests ( $p = 2$, $\alpha = 0.05$).*

| Design | Test/$(n,m)$ | (10,15) | (15,10) | (30,50) | (50,30) | (50,50) | (65,70) | (70,65) |
|---|---|---|---|---|---|---|---|---|
| $D_1$ | $Log(TS_{nm})$ | 0.323 | 0.324 | 0.909 | 0.986 | 0.999 | 1 | 1 |
| | Z | 0.041 | 0.053 | 0.279 | 0.242 | 0.293 | 0.338 | 0.345 |
| | JKM | 0.054 | 0.053 | 0.066 | 0.069 | 0.074 | 0.092 | 0.074 |
| | BG | 0.054 | 0.079 | 0.039 | 0.099 | 0.066 | 0.075 | 0.076 |
| $D_2$ | $Log(TS_{nm})$ | 0.1049 | 0.115 | 0.3205 | 0.3483 | 0.507 | 0.692 | 0.691 |
| | Z | 0.0304 | 0.039 | 0.1412 | 0.1424 | 0.2058 | 0.303 | 0.301 |
| | JKM | 0.1002 | 0.099 | 0.1640 | 0.1940 | 0.203 | 0.250 | 0.248 |
| | BG | 0.1057 | 0.114 | 0.1265 | 0.2946 | 0.2924 | 0.368 | 0.402 |
| $D_3$ | $Log(TS_{nm})$ | 0.080 | 0.092 | 0.704 | 0.5951 | 0.885 | 0.996 | 0.986 |
| | Z | 0.0326 | 0.046 | 0.219 | 0.1661 | 0.213 | 0.386 | 0.390 |
| | JKM | 0.0484 | 0.053 | 0.050 | 0.049 | 0.056 | 0.052 | 0.052 |
| | BG | 0.0430 | 0.056 | 0.094 | 0.080 | 0.105 | 0.134 | 0.145 |
| $D_4$ | $Log(TS_{nm})$ | 0.3398 | 0.356 | 0.825 | 0.843 | 0.922 | 0.983 | 0.985 |
| | Z | 0.0944 | 0.088 | 0.701 | 0.704 | 0.855 | 0.965 | 0.961 |
| | JKM | 0.1962 | 0.191 | 0.365 | 0.357 | 0.419 | 0.486 | 0.484 |
| | BG | 0.1814 | 0.189 | 0.404 | 0.376 | 0.454 | 0.609 | 0.537 |
| $D_5$ | $Log(TS_{nm})$ | 0.0719 | 0.067 | 0.1141 | 0.2231 | 0.2711 | 0.4213 | 0.4215 |
| | Z | 0.0208 | 0.016 | 0.1170 | 0.0920 | 0.0890 | 0.1373 | 0.1365 |
| | JKM | 0.0498 | 0.0493 | 0.0560 | 0.0522 | 0.0523 | 0.0522 | 0.0525 |
| | BG | 0.0386 | 0.063 | 0.0593 | 0.0673 | 0.0641 | 0.0562 | 0.0613 |
| $D_6$ | $Log(TS_{nm})$ | 0.307 | 0.298 | 0.779 | 0.812 | 0.924 | 0.975 | 0.9791 |
| | Z | 0.1054 | 0.064 | 0.601 | 0.591 | 0.758 | 0.888 | 0.9011 |
| | JKM | 0.1631 | 0.154 | 0.323 | 0.366 | 0.385 | 0.415 | 0.4202 |
| | BG | 0.1800 | 0.188 | 0.355 | 0.377 | 0.434 | 0.547 | 0.5282 |
| $D_7$ | $Log(TS_{nm})$ | 0.3204 | 0.319 | 0.78 | 0.788 | 0.927 | 0.9824 | 0.9934 |
| | Z | 0.1022 | 0.084 | 0.665 | 0.619 | 0.783 | 0.9222 | 0.9391 |
| | JKM | 0.1946 | 0.181 | 0.347 | 0.355 | 0.39 | 0.4092 | 0.4224 |
| | BG | 0.2228 | 0.222 | 0.326 | 0.378 | 0.43 | 0.5011 | 0.4930 |
| $D_8$ | $Log(TS_{nm})$ | 0.292 | 0.365 | 0.676 | 0.811 | 0.977 | 1 | 1 |
| | Z | 0.1340 | 0.1132 | 0.7671 | 0.7111 | 0.8911 | 0.9772 | 0.9811 |
| | JKM | 0.1251 | 0.1367 | 0.2576 | 0.2615 | 0.2975 | 0.3377 | 0.3429 |
| | BG | 0.1211 | 0.1422 | 0.1771 | 0.2421 | 0.2881 | 0.3581 | 0.3473 |
| $D_9$ | $Log(TS_{nm})$ | 0.2286 | 0.224 | 0.611 | 0.65 | 0.772 | 0.911 | 0.9195 |
| | Z | 0.0608 | 0.054 | 0.352 | 0.412 | 0.527 | 0.717 | 0.6851 |
| | JKM | 0.1132 | 0.108 | 0.255 | 0.225 | 0.273 | 0.348 | 0.3272 |
| | BG | 0.1371 | 0.137 | 0.186 | 0.189 | 0.255 | 0.306 | 0.3104 |

It seems that the Z, JKM and BG tests can be biased under several scenarios, e.g., $D_5$ with *(n,m)*=(10,15). The $H_1$-design $D_1$ considered samples from a *t*-distribution versus samples from a normal distribution. In this case, the power of the proposed test is roughly three times larger than that of the Z, JKM and BG tests. Note that, the JKM test is developed for location/scale alternatives that are exemplified by the designs $D_4$ and $D_5$. Under $D_4$, $D_5$ and $D_6$ it is clear that the proposed procedure dramatically outperforms the JKM test. $D_8$ and $D_9$ present cases when $X_1, X_2, Y_1, Y_2$ are identically distributed, but $\mathbf{X}$ and $\mathbf{Y}$ have different distributions. Under $D_8$ and $D_9$, the DBEL test is significantly superior to the Z, JKM and BG tests.

In this section we also demonstrate the Monte Carlo power evaluations of the *Log(TS$_{nm}$)*, Z, JKM and BG tests under the following alternative designs, where $p=3$. $S_1$: $\mathbf{X} \sim N(\boldsymbol{\mu}, \Sigma_X)$ vs., $\mathbf{Y} \sim N((0.5,0.7,0.5)^T, \Sigma_X)$, where $\boldsymbol{\mu}=(0,0,0)^T$, $\Sigma_X = \{\sigma_{ij}^2\}$, $\sigma_{ij}^2 = I(i=j), 1 \le i,j \le 3$; $S_2$: $\mathbf{X} \sim N(\boldsymbol{\mu}, \Sigma_X)$ vs. $\mathbf{Y} \sim N(\boldsymbol{\mu}, \Sigma_Y)$, where $\Sigma_Y = \{\tilde{\sigma}_{ij}^2\}$, $\tilde{\sigma}_{ij}^2 = I(i=j) + 0.5I(i \ne j), 1 \le i,j \le 3$; $S_3$: $\mathbf{X} \sim N(\boldsymbol{\mu}, \Sigma_X)$ vs. $\mathbf{Y} \sim N((0.5,0.7,0.5)^T, \Sigma_Y)$, where $\boldsymbol{\mu}, \Sigma_X, \Sigma_Y$ are defined above. Table 5 shows the results of the power evaluations, supporting the experimental conclusions based on the outputs of Table 4.

**Table 5.** *The Monte Carlo power of the tests ($p=3$, $\alpha=0.05$).*

| Design | Test/*(n,m)* | (10,15) | (15,10) | (30,50) | (50,30) | (50,50) |
|---|---|---|---|---|---|---|
| $S_1$ | *Log(TS$_{nm}$)* | 0.3453 | 0.3570 | 0.8338 | 0.8481 | 0.9405 |
| | Z | 0.1215 | 0.0532 | 0.7781 | 0.7235 | 0.9111 |
| | JKM | 0.1865 | 0.1921 | 0.3905 | 0.3713 | 0.3985 |
| | BG | 0.1920 | 0.1845 | 0.4161 | 0.4335 | 0.5471 |
| $S_2$ | *Log(TS$_{nm}$)* | 0.0758 | 0.0701 | 0.16107 | 0.2794 | 0.3251 |
| | Z | 0.0282 | 0.0052 | 0.1330 | 0.1303 | 0.1935 |
| | JKM | 0.0495 | 0.0487 | 0.0495 | 0.0475 | 0.0511 |
| | BG | 0.0501 | 0.0735 | 0.0631 | 0.0775 | 0.0635 |
| $S_3$ | *Log(TS$_{nm}$)* | 0.3547 | 0.345 | 0.8244 | 0.8322 | 0.9354 |

| | | | | | |
|---|---|---|---|---|---|
| Z | 0.0971 | 0.0301 | 0.5115 | 0.5010 | 0.6995 |
| JKM | 0.1510 | 0.1341 | 0.2962 | 0.2890 | 0.3165 |
| BG | 0.1683 | 0.1995 | 0.3845 | 0.3995 | 0.5205 |

Based on the Monte Carlo results, we conclude that the proposed test exhibits very high and stable power characteristics in comparison to the known modern procedures.

## 6. Data Analysis

In this section, we present a data example to illustrate the practical application of the proposed method.

The use of thiobarbituric acid-reactive substances (TBARS) as a value to summarize total circulating oxidative stress in individuals is common in laboratory research (Armstrong, 1994), but its use as a discriminant factor between individuals with and without myocardial infarction (MI) disease is still controversial (e.g., Schisterman *et al*., 2001). Some authors have found a positive association between TBARS and MI disease (e.g., Jayakumari *et al*., 1992; Miwa *et al*., 1995), while others did not find corresponding significant associations (e.g., Karmansky *et al*., 1996). The biomarkers, HDL-cholesterol, glucose and vitamin E, are historically known to be significantly associated with MI disease (e.g., Schisterman *et al*., 2001).

The aim of our study is to investigate the joint discriminative properties of the vector biomarker [TBARS, HDL, glucose, vitamin E]$^T$, with regard to MI disease. Towards this end, we implemented the tests: $Log(TS_{nm})$, Z, JKM and BG described in Section 5, using the following data.

A sample of randomly selected residents of Erie and Niagara counties, 35 to 79 years of age, was employed in this investigation. The New York State department of Motor Vehicles drivers' license rolls was used as the sampling frame for adults between the age of 35 and 65, while the elderly sample (age 65 to 79) was randomly selected from the Health Care Financing Administration database. The study evaluated 70 measurements of TBARS, HDL-cholesterol, glucose and vitamin E biomarkers. Half of them were collected on cases, who recently survived on MI disease (say MI=1), and the other half on controls, who had no previous MI disease (say MI=0). The $p$-values obtained via the $Log(TS_{nm})$, Z, JKM and BG procedures are 0.0050, 0.0579, 0.0501 and 0.0235, respectively. The JKM test provides $p$-values that is slightly larger than a significance level of 5%. The proposed test reveals a strong evidence of an association between a joint distribution of the vector [TBARS, HDL, glucose, vitamin E]$^T$ and MI disease.

Then, we organized a Bootstrap/Jackknife type study to examine the power performances of the test-statistics. The conducted strategy was that two samples with sizes $n = m = 15$ were randomly selected from the data with MI=1 and MI=0 to be tested for the hypothesis $H_0$: the vector of biomarkers values, [TBARS, HDL, glucose, vitamin E]$^T$, is distributed identically with respect to MI=0 and MI=1. The $Log(TS_{nm})$, Z, JKM and BG tests were conducted at 5% level of significance. We repeated this strategy 5,000 times calculating the frequencies of the events {$Log(TS_{nm})$ rejects $H_0$}, {Z rejects $H_0$}, {JKM rejects $H_0$} and {BG rejects $H_0$}. The obtained experimental powers of the four tests are 0.77, 0.19, 0.51, 0.65, respectively. In this study, the proposed test outperforms the known Z, JKM and BG procedures in terms of the power properties when detecting the joint discriminative properties of the biomarkers values, [TBARS, HDL, glucose, vitamin E], with regard to MI disease.

## 7. Concluding Remarks

In this article we developed a novel density-based empirical likelihood ratio mechanism for testing the equality of two multivariate distributions based on observed vectors. The proposed approach is exact and distribution-free.

Our method employs a univariate projection pursuit-based procedure. It is indicated that correctness of applications of projection pursuit are critically important for constructing and performing multivariate decision-making procedures. We proved one-to-one-mapping between the equality of vectors' distributions and the equality of distributions of relevant univariate linear projections. It turns out that an algorithm can be provided to simplify the use of projection pursuit via using only a few of the infinitely many linear combinations of observed vectors' components. In this framework, the demonstrated proof algorithm can be applied to execute different multivariate procedures developed by using Kolmogorov Smirnov type testing mechanisms or ranks based concepts, when the univariate projection pursuit is employed.

The displayed testing strategy was presented in retrospective and group sequential manners. The asymptotic consistency of the proposed technique was shown. Through extensive Monte Carlo experimental studies, we demonstrated that the proposed procedure has significantly higher power as compared with the modern methods of Jurečková and Kalina (2012), Marozzi (2016), Zhou et al. (2017) and Biswas and Ghosh (2014) across a variety of experiments' scenarios, including, e.g., cases when observed $\mathbf{X}$'s and $\mathbf{Y}$'s components are identically distributed, whereas $\mathbf{X}$ and $\mathbf{Y}$ have different distributions. This study shows that the proposed test can efficiently detect relatively small departures from the null hypothesis that treats two multivariate distributions to be identical.

**Supplementary Materials**

**Theoretical Results:** Proofs of Propositions 1,3,4,5 and the algorithm contained in Section 2 regarding the outcome $TS_{nm} = \max_{(u_1,u_2)\in A_1\cup A_2} TS_{nm}\left(X(\mathbf{u}),Y(\mathbf{u})\right)$ $= \max_{(u_1,u_2)\in B_1\cup A_2} TS_{nm}\left(X(\mathbf{u}),Y(\mathbf{u})\right)$; an illustration of the statement of Proposition 3, when $p=3$.

**R Codes:** Codes for Monte Carlo computing the critical values of the null distributions of the proposed tests.

# Supplementary materials to Exact Multivariate Two-Sample Density-Based Empirical Likelihood Ratio Tests Applicable to Retrospective and Group Sequential Studies

Ablert Vexler[1], Gregory Gurevich[2] and Li Zou[3]

[1]*Department of Biostatistics, The State University of New York at Buffalo, Buffalo, NY 14214, U.S.A, avexler@buffalo.edu*

[2]*Department of Industrial Engineering and Management, SCE- Shamoon College of Engineering*

[3]Department of Statistics and Biostatistics, California State University, East Bay, Hayward, CA 94542, U.S.A.

**Supplementary Materials:**

- Proofs of Proposition 1
- To Section 2: The case with $p=2$, $TS_{nm}=\max_{(u_1,u_2)\in A_1\cup A_2} TS_{nm}\left(X(\mathbf{u}),Y(\mathbf{u})\right)$

  $=\max_{(u_1,u_2)\in B_1\cup A_2} TS_{nm}\left(X(\mathbf{u}),Y(\mathbf{u})\right)$.
- Proof of Proposition 3.
- Notations related to Proposition 3's statement with $p=3$.
- Proof of Proposition 4.
- R code to calculate the critical values of the proposed test (see Section 3.1)
- R code to calculate the critical values of the proposed test (see Section 4.1)
- References

## Proof of Proposition 1.

Assume $\Pr\{X(\mathbf{u})\le z\}=\Pr\{Y(\mathbf{u})\le z\}$, for all $z,u_1,...,u_p\in R^1$. Then the characteristic function of $\mathbf{X}$, $\varphi_X\left(t_1,...,t_p\right)=\mathrm{E}\exp\left(i\sum_{j=1}^{p}t_jX_j\right)$ with $t_1,...,t_p\in R^1$, satisfies

$$\varphi_X\left(t_1,...,t_p\right)=\int\exp\left(iz\right)d\Pr\left(\sum_{j=1}^{p}t_jX_j\le z\right)=\int\exp\left(iz\right)d\Pr\left(\sum_{j=1}^{m}t_jY_j\le u\right)=\varphi_Y\left(t_1,...,t_p\right),$$

where $i^2=-1$ and $\varphi_Y\left(t_1,...,t_p\right)$ is the characteristic function of $\mathbf{Y}$. This implies $F_X=F_Y$.

Assume $F_X=F_Y$. We lose nothing in generality if we suppose that there are density functions $f_X$ and $f_Y$ of $\mathbf{X}$ and $\mathbf{Y}$, respectively. The characteristic function of $X(\mathbf{u})$ can be presented as

$$\varphi_{X(u)}(t)=\mathrm{E}\exp\left(it\sum_{j=1}^{p}u_jX_j\right)=\int...\int\exp\left(it\sum_{j=1}^{p}u_jx_j\right)\frac{d...dF_X(x_1,...,x_p)}{dx_1...dx_p}dx_1...dx_p$$

$$=\int...\int\exp\left(it\sum_{j=1}^{p}u_jx_j\right)\frac{d...dF_Y(x_1,...,x_p)}{dx_1...dx_p}dx_1...dx_p=\varphi_{Y(u)}(t),$$

where $t\in R^1$ and $\varphi_{Y(u)}(t)$ is the characteristic function of $Y(\mathbf{u})$. Then we conclude that $X(\mathbf{u})$ and $Y(\mathbf{u})$ are identically distributed for all $u_1,...,u_p\in R^1$.

The proof of Proposition 1 is complete.

## Section 2: The case with $p=2$, $TS_{nm}=\max_{(u_1,u_2)\in A_1\cup A_2}TS_{nm}\left(X(\mathbf{u}),Y(\mathbf{u})\right)$ $=\max_{(u_1,u_2)\in B_1\cup A_2}TS_{nm}\left(X(\mathbf{u}),Y(\mathbf{u})\right)$.

To establish this result, we will apply the following scheme. The statistic $TS_{nm}\left(X(\mathbf{u}),Y(\mathbf{u})\right)$ is based on the variables $I\left\{X_i(\mathbf{u})\le Y_{(j)}(\mathbf{u})\right\}$, $I\left\{Y_j(\mathbf{u})\le X_{(i)}(\mathbf{u})\right\}$, $I\left\{X_i(\mathbf{u})\le X_{(k)}(\mathbf{u})\right\}$, $I\left\{Y_j(\mathbf{u})\le Y_{(r)}(\mathbf{u})\right\}$, $1\le i,k\le n,\ 1\le j,r\le m$. Thus, two vectors $\mathbf{u}=(u_1,u_2)^T$, $\mathbf{v}=(v_1,v_2)^T$ satisfy $TS_{nm}\left(X(\mathbf{u}),Y(\mathbf{u})\right)=TS_{nm}\left(X(\mathbf{v}),Y(\mathbf{v})\right)$, if $I\left\{X_i(\mathbf{u})\le Y_j(\mathbf{u})\right\}=I\left\{X_i(\mathbf{v})\le Y_j(\mathbf{v})\right\}$, $I\left\{X_i(\mathbf{u})\le X_k(\mathbf{u})\right\}=I\left\{X_i(\mathbf{v})\le X_k(\mathbf{v})\right\}$, $I\left\{Y_j(\mathbf{u})\le Y_r(\mathbf{u})\right\}=I\left\{Y_j(\mathbf{v})\le Y_r(\mathbf{v})\right\}$, for all $1\le i,k\le n$ and $1\le j,r\le m$.

Let us focus on the equation $I\left\{X_i(\mathbf{u})\le Y_j(\mathbf{u})\right\}=I\left\{X_i(\mathbf{v})\le Y_j(\mathbf{v})\right\}$. We assume $\mathbf{u}^T=(u_1,u_2)\in A_1$ and consider three scenarios related to locations of values of $u_2$ with respect to the $\{W(i,j)$, $U(i,r)$, $V(j,s)\}$-based order statistics $Q_{(1)}\le Q_{(2)}\le...\le Q_{(\tau)}$, where $\tau=1+2+...+n-1+nm$ $+1+2+...+m-1=(n+m)(n+m-1)/2$.

(1) Suppose that $u_2<Q_{(1)}$, then $u_2<\left(X_{1i}-Y_{1j}\right)\left(Y_{2j}-X_{2i}\right)^{-1}$, for all $1\le i\le n$ and $1\le j\le m$. In this case, since $I\left\{X_{1i}-Y_{1j}>u_2\left(Y_{2j}-X_{2i}\right)\right\}=I\left\{\left(X_{1i}-Y_{1j}\right)\left(Y_{2j}-X_{2i}\right)^{-1}>u_2,\left(Y_{2j}-X_{2i}\right)>0\right\}$

$+I\left\{\left(X_{1i}-Y_{1j}\right)\left(Y_{2j}-X_{2i}\right)^{-1}<u_2,\left(Y_{2j}-X_{2i}\right)\le 0\right\}$, we have $I\left\{X_{1i}-Y_{1j}>u_2\left(Y_{2j}-X_{2i}\right)\right\}$

$=I\left(Y_{2j}-X_{2i}>0\right)$, i.e. $I\left(X_{1i}+u_2X_{2i}>Y_{1j}+u_2Y_{2j}\right)=I\left(\nu_1X_{1i}+\nu_2X_{2i}>\nu_1Y_{1i}+\nu_2Y_{2i}\right)$, where

$\mathbf{v}^T=\left(\nu_1,\nu_2\right)=\left(0,-1\right)$. This implies $TS_{nm}\left(X(\mathbf{u}),Y(\mathbf{u})\right)=TS_{nm}\left(X(\mathbf{v}),Y(\mathbf{v})\right)$. The scenario with $\mathbf{v}^T=\left(0,-1\right)$ corresponds to testing that $-X_{21}$ and $-Y_{21}$ are identically distributed, which is equivalent to testing the distributions of $X_{21}$ and $Y_{21}$. Thus, according to the form of the DBEL ratio test statistic, we have $TS_{nm}\left(X(\mathbf{v}),Y(\mathbf{v})\right)=TS_{nm}\left(X(\mathbf{v}_1),Y(\mathbf{v}_1)\right)$, where $\mathbf{v}_1^T=\left(0,1\right)\in A_2$.

(2) Now, we suppose that $Q_{(d)}\le u_2<Q_{(d+1)}$, $d\in\left\{1,...,\tau-1\right\}$. In this case, the event $\left\{\left(X_{1i}-Y_{1j}\right)\left(Y_{2j}-X_{2i}\right)^{-1}\le u_2\right\}$ equals to the event $\left\{\left(X_{1i}-Y_{1j}\right)\left(Y_{2j}-X_{2i}\right)^{-1}\le Q_{(d)}\right\}$ and then the events $\left\{\left(X_{1i}-Y_{1j}\right)\left(Y_{2j}-X_{2i}\right)^{-1}>u_2\right\}$ and $\left\{\left(X_{1i}-Y_{1j}\right)\left(Y_{2j}-X_{2i}\right)^{-1}>Q_{(d)}\right\}$ are equivalent. This leads to $I\left\{\left(X_{1i}-Y_{1j}\right)\le u_2\left(Y_{2j}-X_{2i}\right)\right\}=I\left\{\left(X_{1i}-Y_{1j}\right)\le u_2\left(Y_{2j}-X_{2i}\right),\left(Y_{2j}-X_{2i}\right)\ge 0\right\}$

$+I\left\{\left(X_{1i}-Y_{1j}\right)\le u_2\left(Y_{2j}-X_{2i}\right),\left(Y_{2j}-X_{2i}\right)<0\right\}=I\left\{\left(X_{1i}-Y_{1j}\right)\le Q_{(d)}\left(Y_{2j}-X_{2i}\right)\right\}$, for all $i=1,...,n,\ j=1,...,m$. This implies $I\left(X_{1i}+u_2X_{2i}\le Y_{1j}+u_2Y_{2j}\right)=I\left(X_{1i}+Q_{(d)}X_{2i}\le Y_{1j}+Q_{(d)}Y_{2j}\right)$, that is $TS_{nm}\left(X(\mathbf{u}),Y(\mathbf{u})\right)=TS_{nm}\left(X(\mathbf{v}),Y(\mathbf{v})\right)$, where $\mathbf{v}^T=\left(0,Q_{(d)}\right)\in B_1$.

(3) Suppose that $u_2>Q_{(\tau)}$, i.e. $\left(X_{1i}-Y_{1j}\right)\left(Y_{2j}-X_{2i}\right)^{-1}<u_2$, for all $i=1,...,n$, $j=1,...,m$. In this case, $I\left\{X_{1i}-Y_{1j}<u_2\left(Y_{2j}-X_{2i}\right)\right\}=I\left\{X_{1i}-Y_{1j}<u_2\left(Y_{2j}-X_{2i}\right),\left(Y_{2j}-X_{2i}\right)\ge 0\right\}$

$+I\left\{X_{1i}-Y_{1j}<u_2\left(Y_{2j}-X_{2i}\right),\left(Y_{2j}-X_{2i}\right)<0\right\}=I\left\{\left(X_{1i}-Y_{1j}\right)\left(Y_{2j}-X_{2i}\right)^{-1}<u_2,\left(Y_{2j}-X_{2i}\right)\ge 0\right\}$

$+I\left\{\left(X_{1i}-Y_{1j}\right)\left(Y_{2j}-X_{2i}\right)^{-1}>u_2,\left(Y_{2j}-X_{2i}\right)<0\right\}=I\left\{\left(Y_{2j}-X_{2i}\right)\ge 0\right\}$. This can be rewritten as

$I\left(X_{1i}+u_2X_{2i}<Y_{1j}+u_2Y_{2j}\right)=I\left(\nu_1X_{1i}+\nu_2X_{2i}<\nu_1Y_{2j}+\nu_2Y_{2j}\right)$ with $\mathbf{v}^T=\left(\nu_1,\nu_2\right)=\left(0,1\right)$ that leads to $TS_{nm}\left(X(\mathbf{u}),Y(\mathbf{u})\right)=TS_{nm}\left(X(\mathbf{v}),Y(\mathbf{v})\right)$, where $\mathbf{v}^T=(0,1)\in A_2$. It is clear that, when $u_2=Q_{(\tau)}$, we have $TS_{nm}\left(X(\mathbf{u}),Y(\mathbf{u})\right)=TS_{nm}\left(X(\mathbf{v}),Y(\mathbf{v})\right)$, where $\mathbf{v}^T=\left(0,Q_{(\tau)}\right)\in B_1$.

The concept outlined above can be easily applied to evaluate the equations $I\left\{X_i(\mathbf{u})\le X_k(\mathbf{u})\right\}=I\left\{X_i(\mathbf{v})\le X_k(\mathbf{v})\right\}$, $I\left\{Y_j(\mathbf{u})\le Y_r(\mathbf{u})\right\}=I\left\{Y_j(\mathbf{v})\le Y_s(\mathbf{v})\right\}$, where $\mathbf{u}\in A_1\cup A_2, \mathbf{v}\in B_1\cup A_2$.

Therefore, in order to compute $TS_{nm}$, we can treat the $\tau$ linear combinations, $X(\mathbf{v}),Y(\mathbf{v})$, with $(v_1,v_2)\in B_1\cup A_2$ only.

## Proof of Proposition 3.

For key principles used in the proof below, we refer the reader to the analysis presented with respect to the case with *p=2.* These principles will be recursively employed to prove Proposition 3.

We will establish that for each vector $\mathbf{u}=\left(u_1,...,u_p\right)^T\in\bigcup_{j=1}^p A_j$, there exists some vector $\mathbf{v}=\left(v_1,...,v_p\right)^T\in\bigcup_{j=1}^p B_j$ such that $TS_{mn}\left(X(\mathbf{u}),Y(\mathbf{u})\right)=TS_{mn}\left(X(\mathbf{v}),Y(\mathbf{v})\right)$. To this end, we consider the indicators $I\left\{X_i(\mathbf{u})\le Y_j(\mathbf{u})\right\}$, $I\left\{X_i(\mathbf{u})\le X_j(\mathbf{u})\right\}$, $I\left\{Y_i(\mathbf{u})\le Y_j(\mathbf{u})\right\}$, $I\left\{X_i(\mathbf{v})\le Y_j(\mathbf{v})\right\}$, $I\left\{X_i(\mathbf{v})\le X_j(\mathbf{v})\right\}$, $I\left\{Y_i(\mathbf{v})\le Y_j(\mathbf{v})\right\}$, on which the statistics $TS_{mn}\left(X(\mathbf{u}),Y(\mathbf{u})\right)$ and $TS_{mn}\left(X(\mathbf{v}),Y(\mathbf{v})\right)$ are based.

We begin with $\mathbf{u}=\left(u_1,...,u_p\right)^T\in A_1$. Denote the order statistics $Q_{2(1)}\le Q_{2(2)}\le...\le Q_{2(\tau_p)}$ based on the set of the variables $W_1\left(J_{W,p-1}\right),U_1\left(J_{U,p-1}\right),V_1\left(J_{V,p-1}\right)$. In this case, $\tau_p$ can be calculated via the recursive formula $\tau_h=\tau_{h-1}\left(\tau_{h-1}-1\right)/2$ with $h=3,...,p$ and $\tau_2=(n+m)(n+m-1)/2$.

We assume that $Q_{2(d_2)}\le u_2<Q_{2(d_2+1)}$, for some $d_2\in\left\{1,...,\tau_p\right\}$. In this case,

$$I\left\{\frac{W_1\left(J_{W,p-2}\right)-W_1\left(J_{W,p-2}^c\right)}{W_2\left(J_{W,p-2}^c\right)-W_2\left(J_{W,p-2}\right)}\le u_2\right\}=I\left\{\frac{W_1\left(J_{W,p-2}\right)-W_1\left(J_{W,p-2}^c\right)}{W_2\left(J_{W,p-2}^c\right)-W_2\left(J_{W,p-2}\right)}\le Q_{2(d_2)}\right\}$$

and then

$$I\left\{W_1\left(J_{W,p-2}\right)-W_1\left(J_{W,p-2}^c\right)\le u_2\left(W_2\left(J_{W,p-2}^c\right)-W_2\left(J_{W,p-2}\right)\right)\right\}$$
$$=I\left\{W_1\left(J_{W,p-2}\right)-W_1\left(J_{W,p-2}^c\right)\le Q_{2(d_2)}\left(W_2\left(J_{W,p-2}^c\right)-W_2\left(J_{W,p-2}\right)\right)\right\},$$

$$I\left\{W_1\left(J_{W,p-2}\right)+W_2\left(J_{W,p-2}\right)u_2\le W_1\left(J_{W,p-2}^c\right)+W_2\left(J_{W,p-2}^c\right)u_2\right\}$$
$$=I\left\{{}^pW_1\left(J_{W,p-2}\right)+W_2\left(J_{W,p-2}\right)Q_{1(d_2)}\le W_1\left(J_{W,p-2}^c\right)+W_2\left(J_{W,p-2}^c\right)Q_{2(d_2)}\right\}.$$

Defining $v_2=Q_{2(d_2)}\in B_1$, we obtain

$$I\left\{W_1\left(J_{W,p-2}\right)+W_2\left(J_{W,p-2}\right)u_2\le W_1\left(J_{W,p-2}^c\right)+W_2\left(J_{W,p-2}^c\right)u_2\right\}$$
$$=I\left\{W_1\left(J_{W,p-2}\right)+W_2\left(J_{W,p-2}\right)v_2\le W_1\left(J_{W,p-2}^c\right)+W_2\left(J_{W,p-2}^c\right)v_2\right\}.$$

This equation means that, for any fixed $J_{W,p-2}$ and fixed $u_2\in\left[Q_{2(d_2)},Q_{2(d_2+1)}\right)$, the rank of $W_1\left(J_{W,p-2}\right)+u_2W_2\left(J_{W,p-2}\right)$ in the sequence of the variables $W_1\left(J_{W,p-2}^c\right)+u_2W_2\left(J_{W,p-2}^c\right)$, for all different $J_{W,p-2}^c$, equals to the rank of $W_1\left(J_{W,p-2}^c\right)+v_2W_2\left(J_{W,p-2}^c\right)$ in the sequence of the variables $W_1\left(J_{W,p-2}^c\right)+v_2W_2\left(J_{W,p-2}^c\right)$, for all different $J_{W,p-2}^c$.

Now we define $O_{3(1)}\le O_{3(2)}\le\ldots\le O_{3\left(\tau_{p-1}\right)}$ to be the order statistics based on $W_1\left(J_{W,p-2}\right)+u_2W_2\left(J_{W,p-2}\right)$, $U_1\left(J_{U,p-2}\right)+u_2U_2\left(J_{U,p-2}\right)$, $V_1\left(J_{V,p-2}\right)+u_2V_2\left(J_{V,p-2}\right)$. Assume that $O_{3(d_3)}\le u_3<O_{3(d_3+1)}$, for some $d_3\in\left\{1,\ldots,\tau_{p-1}\right\}$, then

$$I\left\{W_1\left(J_{W,p-2}\right)+u_2W_2\left(J_{W,p-2}\right)\le u_3\right\}=I\left\{{}^pW_1\left(J_{W,p-2}\right)+v_2W_2\left(J_{W,p-2}\right)\le\tilde{u}_3\right\},$$

for any $\tilde{u}_3$ that satisfies $Q_{3(d_3)}\le\tilde{u}_3<Q_{3(d_3+1)}$, where $Q_{3(1)}\le Q_{3(2)}\le\ldots\le Q_{3\left(\tau_{p-1}\right)}$ are the order statistics based on $W_1\left(J_{W,p-2}\right)+v_2W_2\left(J_{W,p-2}\right)$, $U_1\left(J_{U,p-2}\right)+v_2U_2\left(J_{U,p-2}\right)$, $V_1\left(J_{V,p-2}\right)+v_2V_2\left(J_{V,p-2}\right)$. Thus, we have $I\left\{W_1\left(J_{W,p-2}\right)+u_2W_2\left(J_{W,p-2}\right)\le u_3\right\}=I\left\{W_1\left(J_{W,p-2}\right)+v_2W_2\left(J_{W,p-2}\right)\le Q_{3(d_3)}\right\}$, and denoting $v_3=Q_{3(d_3)}$, we obtain

$$I\left\{W_1\left(J_{W,p-2}\right)+u_2W_2\left(J_{W,p-2}\right)\le u_3\right\}=I\left\{W_1\left(J_{W,p-2}\right)+v_2W_2\left(J_{W,p-2}\right)\le v_3\right\}. \tag{A.1}$$

This leads to

$$I\left\{\frac{W_1\left(J_{W,p-3}\right)-W_1\left(J_{W,p-3}^c\right)}{W_3\left(J_{W,p-3}^c\right)-W_3\left(J_{W,p-3}\right)}+u_2\frac{W_2\left(J_{W,p-3}\right)-W_2\left(J_{W,p-3}^c\right)}{W_3\left(J_{W,p-3}^c\right)-W_3\left(J_{W,p-3}\right)}\le u_3\right\}$$

$$=I\left\{\frac{W_1\left(J_{W,p-3}\right)-W_1\left(J_{W,p-3}^c\right)}{W_3\left(J_{W,p-3}^c\right)-W_3\left(J_{W,p-3}\right)}+v_2\frac{W_2\left(J_{W,p-3}\right)-W_2\left(J_{W,p-3}^c\right)}{W_3\left(J_{W,p-3}^c\right)-W_3\left(J_{W,p-3}\right)}\le v_3\right\},$$

where the definitions of $W_1\left(J_{W,p-2}\right)$ and $W_2\left(J_{W,p-2}\right)$ are employed, that implies

$$I\left\{W_1\left(J_{W,p-3}\right)-W_1\left(J_{W,p-3}^c\right)+u_2\left(W_2\left(J_{W,p-3}\right)-W_2\left(J_{W,p-3}^c\right)\right)\le u_3\left(W_3\left(J_{W,p-3}^c\right)-W_3\left(J_{W,p-3}\right)\right)\right\}$$

$$=I\left\{W_1\left(J_{W,p-3}\right)-W_1\left(J_{W,p-3}^c\right)+v_2\left(W_2\left(J_{W,p-3}\right)-W_2\left(J_{W,p-3}^c\right)\right)\le v_3\left(W_3\left(J_{W,p-3}^c\right)-W_3\left(J_{W,p-3}\right)\right)\right\},$$

and then

$$I\left\{W_1\left(J_{W,p-3}\right)+u_2W_2\left(J_{W,p-3}\right)+u_3W_3\left(J_{W,p-3}\right)\le W_1\left(J_{W,p-3}^c\right)+u_2W_2\left(J_{W,p-3}^c\right)+u_3W_3\left(J_{W,p-3}^c\right)\right\}$$

$$=I\left\{W_1\left(J_{W,p-3}\right)+v_2W_2\left(J_{W,p-3}\right)+v_3W_3\left(J_{W,p-3}\right)\le {}^pW_1\left(J_{W,p-3}^c\right)+v_2W_2\left(J_{W,p-3}^c\right)+v_3W_3\left(J_{W,p-3}^c\right)\right\}.$$

(See the analysis presented regarding the case with *p=2,* for details.)

In a recursive manner, we can employ the scheme shown above in order to obtain that

$$I\left\{\sum_{i=1}^{p-1}u_iW_i\left(J_{W,1}\right)\le\sum_{i=1}^{p-1}u_iW_i\left(J_{W,1}^c\right)\right\}=I\left\{\sum_{i=1}^{p-1}v_iW_i\left(J_{W,1}\right)\le\sum_{i=1}^{p-1}v_iW_i\left(J_{W,1}^c\right)\right\},$$

where $u_1=v_1=1$; it is assumed that $Q_{i(d_i)}\le u_i<Q_{i(d_i+1)}$, for some $d_i\in\{1,...,\tau_{p-i+2}\}$, $i=4,...,p-1$; $O_{i(1)}\le O_{i(2)}\le...\le O_{i(\tau_{p-i+2})}$ are the order statistics based on the variable $\sum_{j=1}^{i-1}u_jW_j\left(J_{W,p-i+1}\right)$, $\sum_{j=1}^{i-1}u_jU_j\left(J_{U,p-i+1}\right)$, $\sum_{j=1}^{i-1}u_jV_j\left(J_{V,p-i+1}\right)$; $v_i=Q_{i(d_i)}$; $Q_{i(1)}\le Q_{i(2)}\le...\le Q_{i(\tau_{p-i+2})}$ are the order statistics based on $\sum_{j=1}^{i-1}v_jW_j\left(J_{W,p-i+1}\right)$, $\sum_{j=1}^{i-1}v_jU_j\left(J_{U,p-i+1}\right)$, $\sum_{j=1}^{i-1}v_jV_j\left(J_{V,p-i+1}\right)$.

In the next step of the proof, we assume that $O_{p(d_p)}\le u_p<O_{p(d_p+1)}$ for $d_p\in\{1,...,\tau_2\}$, where $O_{p(1)} \le O_{p(2)} \le ... \le O_{p(\tau_2)}$ are the order statistics based on $\sum_{j=1}^{p-1} u_j W_j\left(J_{W,1}\right)$, $\sum_{j=1}^{p-1} u_j U_j\left(J_{U,1}\right)$, $\sum_{j=1}^{p-1} u_j V_j\left(J_{V,1}\right)$. Denote $v_p = Q_{p\left(d_p\right)}$, where $Q_{p(1)} \le Q_{p(2)} \le ... \le Q_{p(\tau_2)}$ are the order statistics based on $\sum_{j=1}^{p-1} v_j W_j\left(J_{W,1}\right)$, $\sum_{j=1}^{p-1} v_j U_j\left(J_{U,1}\right)$, $\sum_{j=1}^{p-1} v_j V_j\left(J_{V,1}\right)$. It is clear that we can apply the concept used for obtaining Equation (A.1) to conclude that $I\left\{\sum_{j=1}^{p-1} u_j W_j\left(J_{W,1}\right) \le u_p\right\} = I\left\{\sum_{j=1}^{p-1} v_j W_j\left(J_{W,1}\right) \le v_p\right\}$.

This result can be rewritten as

$$I\left\{\sum_{k=1}^{p-1} u_k\left(X_{ki} - Y_{kj}\right)\left(Y_{pj} - X_{pi}\right)^{-1} \le u_p\right\} = I\left\{\sum_{k=1}^{p-1} v_k\left(X_{ki} - Y_{kj}\right)\left(Y_{pj} - X_{pi}\right)^{-1} \le v_p\right\}.$$

Then,

$I\left(\sum_{k=1}^{p} u_k X_{ki} \le \sum_{k=1}^{p} u_k Y_{ki}\right) = I\left(\sum_{k=1}^{p} v_k X_{ki} \le \sum_{k=1}^{p} v_k Y_{ki}\right)$ for all $i \in [1,...,n], j \in [1,...,m]$.

That is, it turns out that if $\mathbf{u} = \left(u_1,...,u_p\right)^T \in A_1$ when $O_{i(d_i)} \le u_i < O_{i(d_i+1)}$ with $d_i \in \left\{1,...,\tau_{p-i+2}\right\}$, $v_i = Q_{i\left(d_i\right)}, i = 2,..., p$, then $I\left\{X_i(\mathbf{u}) \le Y_j(\mathbf{u})\right\} = I\left\{X_i(\mathbf{v}) \le Y_j(\mathbf{v})\right\}$ for all $i = 1,...,n$ and $j = 1,...,m$, where $\mathbf{v} = \left(v_1,...,v_p\right)^T \in B_1$.

Now, we consider the scenario with $u_p > O_{p(\tau_2)}$. In this case, $\sum_{k=1}^{p-1} u_k W_k\left(J_{W,1}\right) < u_p$, i.e. $\sum_{k=1}^{p-1} u_k\left(X_{ki} - Y_{kj}\right)\left(Y_{pj} - X_{pi}\right)^{-1} < u_p$, for all $i = 1,...,n$ and $j = 1,...,m$. Hence, we have $I\left\{\sum_{k=1}^{p-1} u_k\left(X_{ki} - Y_{kj}\right) < u_p\left(Y_{pj} - X_{pi}\right)\right\} = I\left(Y_{pj} - X_{pi} > 0\right)$, This can be represented as $I\left(\sum_{k=1}^{p} u_k X_{ki} < \sum_{k=1}^{p} u_k Y_{kj}\right) = I\left(\sum_{k=1}^{p} v_k X_{ki} < \sum_{k=1}^{p} v_k Y_{kj}\right)$ with $\mathbf{v}^T = \left(v_1, v_2,..., v_p\right) = (0,...,0,1)$ that leads to $TS_{nm}\left(X(\mathbf{u}), Y(\mathbf{u})\right) = TS_{nm}\left(X(\mathbf{v}), Y(\mathbf{v})\right)$, where $\mathbf{v}^T = (0,...,0,1) \in B_p = A_p$.

In a similar manner to the algorithm shown above, we have that if $\sum_{k=1}^{p-1} u_k W_k\left(J_{W,1}\right) > u_p$, i.e. $\sum_{k=1}^{p-1} u_k\left(X_{ki} - Y_{kj}\right)\left(Y_{pj} - X_{pi}\right)^{-1} > u_p$, for all $i = 1,...,n$ and $j = 1,...,m$. We obtain $I\left\{\sum_{k=1}^{p-1} u_k\left(X_{ki} - Y_{kj}\right) > u_p\left(Y_{pj} - X_{pi}\right)\right\} = I\left(Y_{pj} - X_{pi} > 0\right)$, This implies $I\left(\sum_{k=1}^{p} u_k X_{ki} > \sum_{k=1}^{p} u_k Y_{kj}\right) = I\left(\sum_{k=1}^{p} v_k X_{ki} > \sum_{k=1}^{p} v_k Y_{kj}\right)$ with $\mathbf{v}^T = \left(v_1, v_2, ..., v_p\right) = \left(0, ..., 0, -1\right)$ and then $TS_{nm}\left(X(\mathbf{u}), Y(\mathbf{u})\right) = TS_{nm}\left(X(\mathbf{v}), Y(\mathbf{v})\right)$, where $\mathbf{v}^T = \left(0, ..., 0, -1\right)$. The scenario with $\mathbf{v}^T = \left(0, ..., 0, -1\right)$ corresponds to testing that $-X_{p1}$ and $-Y_{p1}$ are identically distributed, which is equivalent to testing distributions of $X_{p1}$ and $Y_{p1}$. Thus, according to the form of the DBEL ratio test statistic, we have $TS_{nm}\left(X(\mathbf{v}), Y(\mathbf{v})\right) = TS_{nm}\left(X(\mathbf{v}_1), Y(\mathbf{v}_1)\right)$, where $\mathbf{v}_1^T = \left(0, ..., 0, 1\right) \in B_p = A_p$.

We next consider the cases with $O_{i(d_i)} \le u_i < O_{i(d_i+1)}$ and $u_j > O_{j\left(\tau_{p-j+2}\right)}$, where $d_i \in \left\{1, ..., \tau_{p-i+2}\right\}$, $i = j+1, ..., p$ and $j = 2, ..., p-1$. In these cases, $\sum_{k=1}^{j-1} u_k W_k\left(J_{W,p-j+1}\right) < u_j$. Then,

$\sum_{k=1}^{j-1} u_k \left(W_k\left(J_{W,p-j}\right) - W_k\left(J_{W,p-j}^c\right)\right)\left(W_j\left(J_{W,p-j}^c\right) - W_j\left(J_{W,p-j}\right)\right)^{-1} < u_j$. This provides

$I\left\{\sum_{k=1}^{j-1} u_k \left(W_k\left(J_{W,p-j}\right) - W_k\left(J_{W,p-j}^c\right)\right) < u_j \left(W_j\left(J_{W,p-j}^c\right) - W_j\left(J_{W,p-j}\right)\right)\right\}$

$= I\left(W_j\left(J_{W,p-j}^c\right) - W_j\left(J_{W,p-j}\right) > 0\right)$, i.e. $I\left\{\sum_{k=1}^{j} u_k W_k\left(J_{W,p-j}\right) < \sum_{k=1}^{j} u_k W_k\left(J_{W,p-j}^c\right)\right\}$

$= I\left\{W_j\left(J_{W,p-j}\right) < W_j\left(J_{W,p-j}^c\right)\right\}$, for a fixed vector $J_{W,p-j}$ and different $J_{W,p-j}^c$. That is, for a fixed vector $J_{W,p-j}$, the tank of $\sum_{k=1}^{j} u_k W_k\left(J_{W,p-j}\right)$ in the sequence of the variables $\sum_{k=1}^{j} u_k W_k\left(J_{W,p-j}^c\right)$, for all different $J_{W,p-j}^c$, equals to that of $W_j\left(J_{W,p-j}\right)$ in the sequence of $W_j\left(J_{W,p-j}^c\right)$, for all different $J_{W,p-j}^c$. For $u_s \in \left[O_{s\,(d_s)}, O_{s\,(d_s+1)}\right)$, we have

$I\left\{\sum_{k=1}^{j} u_k W_k\left(J_{W,p-j}\right) \le u_s\right\} = I\left\{W_j\left(J_{W,p-j}\right) \le \breve{u}_s\right\}$, where $s = j+1$, $\breve{u}_s \in \left[Z_{(d_s)}, Z_{(d_s+1)}\right]$, $Z_{(1)} \le Z_{(2)} \le ...$ are the order statistics based on $W_j\left(J_{W,p-j}\right)$. Hence

$I\left\{\sum_{k=1}^{j} u_k W_k\left(J_{W,p-j}\right) \le u_{j+1}\right\} = I\left\{W_j\left(J_{W,p-j}\right) \le v_{j+1}\right\}$, where $v_s = Z_{(d_s)}, s = j+1$. By virtue of the definitions of $W_k\left(J_{W,p-j}\right)$, we obtain

$$I\left\{\sum_{k=1}^{j} u_k \frac{W_k\left(J_{W,p-j-1}\right)-W_k\left(J^c_{W,p-j-1}\right)}{W_{j+1}\left(J^c_{W,p-j-1}\right)-W_{j+1}\left(J_{W,p-j-1}\right)} \le u_{j+1}\right\} = I\left\{\frac{W_j\left(J_{W,p-j-1}\right)-W_j\left(J^c_{W,p-j-1}\right)}{W_{j+1}\left(J^c_{W,p-j-1}\right)-W_{j+1}\left(J_{W,p-j-1}\right)} \le v_{j+1}\right\}$$

that gives $I\left\{\sum_{k=1}^{j+1} u_k W_k\left(J_{W,p-j-1}\right) \le \sum_{k=1}^{j+1} u_k W_k\left(J^c_{W,p-j-1}\right)\right\}$

$$= I\left\{W_k\left(J_{W,p-j-1}\right)+v_{j+1}W_{j+1}\left(J_{W,p-j-1}\right) \le W_k\left(J_{W,p-j-1}\right)+v_{j+1}W_{j+1}\left(J^c_{W,p-j-1}\right)\right\}.$$

The same arguments employed in dealing with the proof scheme shown above immediately lead to $I\left\{\sum_{k=1}^{p-1} u_k W_k\left(J_{W,2}\right) \le \sum_{k=1}^{p-1} u_k W_k\left(J^c_{W,2}\right)\right\}$

$= I\left\{W_j\left(J_{W,2}\right)+\sum_{k=j+1}^{p-1} v_k W_k\left(J_{W,2}\right) \le W_j\left(J^c_{W,2}\right)+\sum_{k=j+1}^{p-1} v_k W_k\left(J^c_{W,2}\right)\right\}$ that means, for all $s$,

$I\left(\sum_{k=1}^{p} u_k X_{ki} \le \sum_{k=1}^{p} u_k Y_{ki}\right) = I\left(X_{ji}+\sum_{k=j+1}^{p} v_k X_{ki} \le Y_{js}+\sum_{k=j+1}^{p} v_k Y_{ks}\right)$ with $v_{j+k} = Q_{(d_{j+k})}$, where $Q_{(1)} \le Q_{(2)} \le ....$ are the order statistic based on $W_j\left(J_{W,p-j-k+2}\right)+\sum_{i=j+1}^{j+k-1} v_i W_i\left(J_{W,p-j-k+2}\right)$, $1 \le d_{j+k} \le \tau_{p-j-k+2}$. $k = 2,...,p-j$. That is, $TS_{nm}\left(X(\mathbf{u}),Y(\mathbf{u})\right) = TS_{nm}\left(X(\mathbf{v}),Y(\mathbf{v})\right)$, where $\mathbf{v}^T = \left(0,...,0,1,v_{j+1},...,v_p\right) \in B_j$.

By similar techniques, we can obtain the following results:

If, for all $i = j+1,...,p$, we have $O_{i(d_i)} \le u_i < O_{i(d_i+1)}$ with $d_i \in \left\{1,...,\tau_{p-i+2}\right\}$, and $u_j < O_{j(1)}$, where $2 \le j < p$, then we can find a vector $\mathbf{w}^T = \left(0,...,0,-1,-w_{j+1},...,-w_p\right)$ that satisfies $TS_{nm}\left(X(\mathbf{u}),Y(\mathbf{u})\right) = TS_{nm}\left(X(\mathbf{w}),Y(\mathbf{w})\right)$ and $\mathbf{v}^T = \left(0,...,0,1,w_{j+1},...,w_p\right) \in B_j$. Since testing that random variables $\eta$ and $\xi$ are identically distributed is equivalent to testing equality of distributions of $-\eta$ and $-\xi$, according to the form of the DBEL ratio test statistic, we have $TS_{nm}\left(X(\mathbf{w}),Y(\mathbf{w})\right) = TS_{nm}\left(X(\mathbf{v}),Y(\mathbf{v})\right)$.

It can be shown that if $u_2 < Q_{2(1)}$ or $u_2 > Q_{2(\tau_p)}$ then $TS_{nm}\left(X(\mathbf{u}),Y(\mathbf{u})\right) = TS_{nm}\left(X(\mathbf{v}),Y(\mathbf{v})\right)$, where $\mathbf{v}^T \in B_2$.

The proof scheme displayed above can be directly applied to analyze the equations

$I\{X_i(\mathbf{u}) \le X_r(\mathbf{u})\} = I\{X_i(\mathbf{v}) \le X_r(\mathbf{v})\}$, $I\{Y_j(\mathbf{u}) \le Y_s(\mathbf{u})\} = I\{Y_j(\mathbf{v}) \le Y_s(\mathbf{v})\}$, $1 \le i, r \le n$, $1 \le j, s \le m$ as

well as to consider the situations with $\mathbf{u} = \left(u_1, ..., u_p\right)^T \in A_j$, $j = 2, ..., p-1$.

The proof of Proposition 3 is complete.

## Notations related to Proposition 3's statement with $p = 3$.

$B_1 = \left[\left(u_1, u_2, u_3\right):\right.$

$u_1 = 1; u_2 \in \left\{W_1\left(J_{W,2}\right), U_1\left(J_{U,2}\right), V_1\left(J_{V,2}\right), \text{when } J_{W,1} \ne J_{W,1}^c, J_{U,1} \ne J_{U,1}^c, J_{V,1} \ne J_{V,1}^c\right\}$;

given $u_2$ chosen above, $u_3 \in \left\{W_1\left(J_{W,1}\right) + u_2 W_2\left(J_{W,1}\right), U_1\left(J_{U,1}\right) + u_2 U_2\left(J_{U,1}\right), V_1\left(J_{V,1}\right) + u_2 V_2\left(J_{V,1}\right)\right\}$,

$B_2 = \left[\left(u_1, u_2, u_3\right):\ u_1 = 0; u_2 = 1; u_3 \in \left\{W_2\left(J_{W,1}\right), U_2\left(J_{U,1}\right), V_2\left(J_{V,1}\right)\right\}\right]$,

$B_3 = A_3 = \left\{\left(u_1 = 0, u_2 = 0, u_3 = 1\right)\right\}$,

where we have:

$W_1\left(J_{W,2}\right) = \dfrac{W_1\left(J_{W,1}\right) - W_1\left(J_{W,1}^c\right)}{W_2\left(J_{W,1}^c\right) - W_2\left(J_{W,1}\right)}$, $W_1\left(J_{W,1}\right) = \dfrac{X_{1i_1} - Y_{1j_1}}{Y_{3j_1} - X_{3i_1}}$, $W_1\left(J_{W,1}^c\right) = \dfrac{X_{1i_2} - Y_{1j_2}}{Y_{3j_2} - X_{3i_2}}$,

$W_2\left(J_{W,1}\right) = \dfrac{X_{2i_1} - Y_{2j_1}}{Y_{3j_1} - X_{3i_1}}$ and $W_2\left(J_{W,1}^c\right) = \dfrac{X_{2i_2} - Y_{2j_2}}{Y_{3j_2} - X_{3i_2}}$;

$W_1\left(J_{W,2}\right) = \dfrac{\dfrac{X_{1i_1} - Y_{1j_1}}{Y_{3j_1} - X_{3i_1}} - \dfrac{X_{1i_2} - Y_{1j_2}}{Y_{3j_2} - X_{3i_2}}}{\dfrac{X_{2i_2} - Y_{2j_2}}{Y_{3j_2} - X_{3i_2}} - \dfrac{X_{2i_1} - Y_{2j_1}}{Y_{3j_1} - X_{3i_1}}}$, $i_1, i_2 = 1, ..., n$, $j_1, j_2 = 1, ..., m$, $\left(i_1, j_1\right) \ne \left(i_2, j_2\right)$,

$W_1\left(J_{W,1}\right) + u_2 W_2\left(J_{W,1}\right) = \dfrac{X_{1i_1} - Y_{1j_1}}{Y_{3j_1} - X_{3i_1}} + u_2 \dfrac{X_{2i_1} - Y_{2j_1}}{Y_{3j_1} - X_{3i_1}}$, $i_1 = 1, ..., n$, $j_1 = 1, ..., m$,

$U_1\left(J_{U,2}\right) = \dfrac{U_1\left(J_{U,1}\right) - U_1\left(J_{U,1}^c\right)}{U_2\left(J_{U,1}^c\right) - U_2\left(J_{U,1}\right)}$, $U_1\left(J_{U,1}\right) = \dfrac{X_{1i_1} - X_{1r_1}}{X_{3r_1} - X_{3i_1}}$, $U_1\left(J_{W,1}^c\right) = \dfrac{X_{1i_2} - X_{1r_2}}{X_{3r_2} - X_{3i_2}}$,

$U_2\left(J_{U,1}\right) = \dfrac{X_{2i_1} - X_{2r_1}}{X_{3r_1} - X_{3i_1}}$ and $U_2\left(J_{W,1}^c\right) = \dfrac{X_{2i_2} - X_{2r_2}}{X_{3r_2} - X_{3i_2}}$;

$$U_1\left(J_{U,2}\right)=\frac{\dfrac{X_{1i_1}-X_{1r_1}}{X_{3r_1}-X_{3i_1}}-\dfrac{X_{1i_2}-X_{1r_2}}{X_{3r_2}-X_{3i_2}}}{\dfrac{X_{2i_2}-X_{2r_2}}{X_{3r_2}-X_{3i_2}}-\dfrac{X_{2i_1}-X_{2r_1}}{X_{3r_1}-X_{3i_1}}},\ i_1,i_2,r_1,r_2=1,...,n,\ \left(i_1,r_1\right)\neq\left(i_2,r_2\right),\ i_j\neq r_j,\ j=1,2,$$

$$U_1\left(J_{U,1}\right)+u_2U_2\left(J_{U,1}\right)=\frac{X_{1i_1}-X_{1r_1}}{X_{3r_1}-X_{3i_1}}+u_2\frac{X_{2i_1}-X_{2r_1}}{X_{3r_1}-X_{3i_1}},\ i_1,r_1=1,...,n,\ i_1\neq r_1;$$

$$V_1\left(J_{V,2}\right)=\frac{{}^pV_1\left(J_{V,1}\right)-{}^pV_1\left(J_{V,1}^c\right)}{{}^pV_2\left(J_{V,1}^c\right)-{}^pV_2\left(J_{V,1}\right)},\qquad V_1\left(J_{W,1}\right)=\frac{Y_{1j_1}-Y_{1s_1}}{Y_{3s_1}-X_{3j_1}},\qquad V_1\left(J_{V,1}^c\right)=\frac{Y_{1j_2}-Y_{1s_2}}{Y_{3s_2}-Y_{3j_2}},$$

$$V_2\left(J_{V,1}\right)=\frac{Y_{2j_1}-Y_{2s_1}}{Y_{3s_1}-X_{3j_1}},\ V_2\left(J_{V,1}^c\right)=\frac{Y_{2j_2}-Y_{2s_2}}{Y_{3s_2}-Y_{3j_2}};$$

$$V_1\left(J_{V,2}\right)=\frac{\dfrac{Y_{1j_1}-Y_{1s_1}}{Y_{3s_1}-Y_{3j_1}}-\dfrac{Y_{1j_2}-Y_{1s_2}}{Y_{3s_2}-Y_{3j_2}}}{\dfrac{Y_{2j_2}-Y_{2s_2}}{Y_{3s_2}-Y_{3j_2}}-\dfrac{Y_{2j_1}-Y_{2s_1}}{Y_{3s_1}-Y_{3j_1}}},\ j_1,j_2,s_1,s_2=1,...,m,\ \left(j_1,s_1\right)\neq\left(j_2,s_2\right),\ j_i\neq s_i,\ i=1,2,$$

$$V_1\left(J_{V,1}\right)+u_2V_2\left(J_{V,1}\right)=\frac{Y_{1j_1}-Y_{1s_1}}{Y_{3s_1}-Y_{3j_1}}+u_2\frac{Y_{2j_1}-Y_{2s_1}}{Y_{3s_1}-Y_{3j_1}},\ j_1,s_1=1,...,m,\ j_1\neq s_1.$$

## Proof of Proposition 4.

In order to simplify the needed proof scheme of Proposition 3, it is clear that, we can focus on the case with bivariate observations and, by virtue of Proposition 2, we can state the problem in the following form. Let $X_i(a)=X_{1i}+aX_{2i}, i=1,...,n,$ and $Y_i(a)=Y_{1i}+aY_{2i}, i=1,...,m,$ where $a\in R^1$. Assume that $X_i(a)\sim f_{H_0}(u;a)$, under $H_0$; $X_i(a)\sim f_{X(a)}(u;a)$, under $H_1$; $Y_i(a)\sim f_{H_0}(u;a)$, under $H_0$; $Y_i(a)\sim f_{Y(a)}(u;a)$, under $H_1$. Let a distribution function $F_{H_0}(u;a)$ correspond to the $H_0$-density $f_{H_0}(u;a)$. Define the DBEL ratios

$$ELR_{X,n}(a)=\min_{a_n\le k\le b_n}\prod_{i=1}^{n}\frac{2k}{n\left\{F_{n+k}\left(X_{(i+k)}(a)\right)-F_{n+k}\left(X_{(i-k)}(a)\right)\right\}},$$

$$ELR_{Y,m}(a)=\min_{a_m\le r\le b_m}\prod_{i=1}^{m}\frac{2r}{m\left\{F_{n+k}\left(Y_{(i+r)}(a)\right)-F_{n+k}\left(Y_{(i-r)}(a)\right)\right\}},\ a_j=j^{0.5+\delta},\ b_j=\min\left(j^{1-\delta},0.5j\right),$$

$j=n,m$, $\delta\in(0,0.25)$, where $F_{n+m}(t)=(n+m)^{-1}\left\{\sum_{i=1}^{n}I\left(X_i(a)\le t\right)+\sum_{j=1}^{m}I\left(Y_j(a)\le t\right)\right\}$ is the $H_0$-empirical distribution function, $X_{(1)}(a)\le X_{(2)}(a)\le ...\le X_{(n)}(a)$ and $Y_{(1)}(a)\le Y_{(2)}(a)\le ...\le Y_{(m)}(a)$ are the order statistics based on observations $X_i(a), i=1,...,n$ and $Y_i(a), i=1,...,m$, respectively. Here $X_{(i+r)}(a)=X_{(n)}(a)$, if $i+r>n$; $X_{(i-r)}(a)=X_{(1)}(a)$, if $i-r<1$; $Y_{(i+s)}(a)=Y_{(m)}(a)$, if $i+s>m$; $Y_{(i-s)}(a)=Y_{(1)}(a)$, if $i-s<1$. Note that, we can present $F_{n+m}(t)=(n+m)^{-1}\left\{nF_{X(a)n}(t)+mF_{Y(a)m}(t)\right\}$, where the empirical distribution functions

$$F_{X(a)n}(t)=n^{-1}\sum_{i=1}^{n}I\left(X_i(a)\le t\right),\ F_{Y(a)m}(t)=m^{-1}\sum_{i=1}^{m}I\left(Y_i(a)\le t\right).$$

We consider the statistic $TS_{nm}=\max_a ELR_{X,n}(a)ELR_{Y,m}(a)$, aiming to prove that there is a positive threshold $C$, such that $\Pr_{H_1}\left\{(n+m)^{-1}\log\left(TS_{nm}\right)>C\right\}\to 1$ and $\Pr_{H_0}\left\{(n+m)^{-1}\log\left(TS_{nm}\right)>C\right\}\to 0$, when $n/m\to\eta>0$ as $n\to\infty$.

Define

$$\gamma(\mathbf{u}_0)=-\frac{\eta}{1+\eta}\mathrm{E}_{H_1}\left[\log\left\{\frac{\eta}{1+\eta}+\frac{1}{1+\eta}\frac{f_{Y(\mathbf{u}_0)}\left(X_1(\mathbf{u}_0);\mathbf{u}_0\right)}{f_{X(\mathbf{u}_0)}\left(X_1(\mathbf{u}_0);\mathbf{u}_0\right)}\right\}\right]$$
$$-\frac{1}{1+\eta}\mathrm{E}_{H_1}\left[\log\left\{\frac{1}{1+\eta}+\frac{\eta}{1+\eta}\frac{f_{X(\mathbf{u}_0)}\left(Y_1(\mathbf{u}_0);\mathbf{u}_0\right)}{f_{Y(\mathbf{u}_0)}\left(Y_1(\mathbf{u}_0);\mathbf{u}_0\right)}\right\}\right]$$

and note that

$$\gamma(\mathbf{u}_0)>-\frac{\eta}{1+\eta}\log\left\{\frac{\eta}{1+\eta}+\frac{1}{1+\eta}\mathrm{E}_{H_1}\frac{f_{Y(\mathbf{u}_0)}\left(X_1(\mathbf{u}_0);\mathbf{u}_0\right)}{f_{X(\mathbf{u}_0)}\left(X_1(\mathbf{u}_0);\mathbf{u}_0\right)}\right\}$$
$$-\frac{1}{1+\eta}\log\left\{\frac{1}{1+\eta}+\frac{\eta}{1+\eta}\mathrm{E}_{H_1}\frac{f_{X(\mathbf{u}_0)}\left(Y_1(\mathbf{u}_0);\mathbf{u}_0\right)}{f_{Y(\mathbf{u}_0)}\left(Y_1(\mathbf{u}_0);\mathbf{u}_0\right)}\right\}=0.$$

**Lemma A.1.** For a fixed $a_0$, $\gamma = \gamma\left(\left(1, a_0\right)^T\right)$ and a threshold $C$, which satisfies $0 < C < \gamma$, we have $\Pr_{H_1}\left\{\left(n+m\right)^{-1}\log\left(TS_{nm}\right) > C\right\} \to 1$, as $n \to \infty$.

**Proof.** $\Pr_{H_1}\left\{\left(n+m\right)^{-1}\log\left(TS_{nm}\right) > C\right\} \geq \Pr_{H_1}\left\{\left(n+m\right)^{-1}\log\left(ELR_{X,n}(a_0)ELR_{Y,m}(a_0)\right) > C\right\} \to 1$ as $n \to \infty$, since Proposition 4.1 of Gurevich and Vexler (2011). This completes the proof of Lemma A.1.

Now, we assume that the null hypothesis is true and obtain the following result.

**Lemma A.2.** Under $H_0$, for all $C > 0$, we have $\Pr_{H_0}\left\{\left(n+m\right)^{-1}\log\left(TS_{nm}\right) > C\right\} \to 0$, as $n \to \infty$.

**Proof.** We begin with the trivial inequality: since the definitions of $ELR_{X,n}(a)$ and $ELR_{Y,m}(a)$ involve the minimums, we have

$$\begin{aligned}
&\Pr_{H_0}\left\{\left(n+m\right)^{-1}\log\left(TS_{nm}\right) > C\right\} \\
&\leq \Pr_{H_0}\left\{\left(n+m\right)^{-1}\max_a \log\left(\prod_{i=1}^{n}\frac{2n^{1-\gamma}}{n}\left\{F_{n+m}\left(X_{\left(i+n^{1-\gamma}\right)}(a)\right) - F_{n+m}\left(X_{\left(i-n^{1-\gamma}\right)}(a)\right)\right\}^{-1}\right.\right. \\
&\qquad \left.\left.\prod_{i=1}^{m}\frac{2m^{1-\gamma}}{m}\left\{F_{n+m}\left(Y_{\left(i+m^{1-\gamma}\right)}(a)\right) - F_{n+m}\left(Y_{\left(i-m^{1-\gamma}\right)}(a)\right)\right\}^{-1}\right) > C\right\},
\end{aligned}$$

where $\gamma < 1/4$. Then, using the definitions of $F_{n+m}\left(t\right) = \left(n+m\right)^{-1}\left\{nF_{X(a)n}(t) + mF_{Y(a)m}(t)\right\}$ and the order statistics $X_{(i+r)}(a), X_{(i-r)}(a), Y_{(i+s)}(a), Y_{(i-r)}(a)$, we can represent

$$\Pr_{H_0}\left\{\left(n+m\right)^{-1}\log\left(TS_{nm}\right) > C\right\} \leq \Pr_{H_0}\left[\left(n+m\right)^{-1}\min_a\left\{A_{1kn}(a) + A_{2mn}(a)\right\} < -C\right]$$

with

$$A_{1mn}(a) = \sum_{i=1}^{n}\log\left[\frac{n}{2n^{1-\gamma}(n+m)}\left\{g_n\left(i+n^{1-\gamma}\right) + mF_{Y(a)m}\left(X_{\left(i+n^{1-\gamma}\right)}(a)\right) - \left(g_n\left(i-n^{1-\gamma}\right) + mF_{Y(a)m}\left(X_{\left(i-n^{1-\gamma}\right)}(a)\right)\right)\right\}\right],$$

$$A_{2mn}(a) = \sum_{j=1}^{m}\log\left[\frac{m}{2m^{1-\gamma}(n+m)}\left\{g_n\left(j+m^{1-\gamma}\right) + nF_{X(a)n}\left(Y_{\left(j+m^{1-\gamma}\right)}(a)\right) - \left(g_n\left(j-m^{1-\gamma}\right) + nF_{X(a)n}\left(Y_{\left(i-m^{1-\gamma}\right)}(a)\right)\right)\right\}\right]$$

and the function $g_n\left(t\right) = I\left(t \leq 0\right) + tI\left(0 < t \leq n\right) + nI(t > n)$.

Now, to replace the empirical distribution functions with the corresponding theoretical distributions, we define

$$G_{1imn}(a)=F_{Y(a)m}\left(X_{\left(i+n^{1-\gamma}\right)}(a)\right)-F_{H_0}\left(X_{\left(i+n^{1-\gamma}\right)}(a);a\right),\ G_{2imn}(a)=F_{H_0}\left(X_{\left(i+n^{1-\gamma}\right)}(a);a\right)-F_{X(a)n}\left(X_{\left(i+n^{1-\gamma}\right)}(a)\right),$$

$$G_{3imn}(a)=F_{H_0}\left(X_{\left(i-n^{1-\gamma}\right)}(a);a\right)-F_{Y(a)m}\left(X_{\left(i-n^{1-\gamma}\right)}(a)\right),\ G_{4imn}(a)=F_{X(a)n}\left(X_{\left(i-n^{1-\gamma}\right)}(a)\right)-F_{H_0}\left(X_{\left(i-n^{1-\gamma}\right)}(a);a\right),$$

$$\Psi_{1jmn}(a)=F_{X(a)n}\left(Y_{\left(j+m^{1-\gamma}\right)}(a)\right)-F_{H_0}\left(Y_{\left(j+m^{1-\gamma}\right)}(a);a\right),\ \Psi_{2jmn}(a)=F_{H_0}\left(Y_{\left(j+m^{1-\gamma}\right)}(a);a\right)-F_{Y(a)m}\left(Y_{\left(j+m^{1-\gamma}\right)}(a)\right),$$

$$\Psi_{3jmn}(a)=F_{H_0}\left(Y_{\left(j-m^{1-\gamma}\right)}(a);a\right)-F_{X(a)n}\left(Y_{\left(j-m^{1-\gamma}\right)}(a)\right),\ \Psi_{4jmn}(a)=F_{Y(a)m}\left(Y_{\left(i-m^{1-\gamma}\right)}(a)\right)-F_{H_0}\left(Y_{\left(i-m^{1-\gamma}\right)}(a);a\right),$$

and rewrite

$$A_{1mn}(a)=\sum_{i=1}^{n}\log\left[\frac{n}{2n^{1-\gamma}(n+m)}\left\{g_n\left(i+n^{1-\gamma}\right)-g_n\left(i-n^{1-\gamma}\right)\right.\right.$$
$$\left.\left.+mF_{X(a)n}\left(X_{\left(i+n^{1-\gamma}\right)}(a)\right)-mF_{X(a)n}\left(X_{\left(i-n^{1-\gamma}\right)}(a)\right)+m\sum_{k=1}^{4}G_{kimn}(a)\right\}\right],$$
$$A_{2mn}(a)=\sum_{j=1}^{m}\log\left[\frac{m}{2m^{1-\gamma}(n+m)}\left\{g_n\left(j+m^{1-\gamma}\right)-g_n\left(j-m^{1-\gamma}\right)\right.\right.$$
$$\left.\left.+nF_{Y(a)m}\left(Y_{\left(j+m^{1-\gamma}\right)}(a)\right)-nF_{Y(a)m}\left(Y_{\left(j-m^{1-\gamma}\right)}(a)\right)+n\sum_{k=1}^{4}\Psi_{kjmn}(a)\right\}\right].$$

Then, we obtain the following inequality. For $\varepsilon>0$,

$$\Pr_{H_0}\left\{(n+m)^{-1}\log\left(TS_{nm}\right)>C\right\}$$
$$\leq\Pr_{H_0}\left[(n+m)^{-1}\min_a\left\{A_{1nm}(a)+A_{2nm}(a)\right\}<-C,\ \max_a G_m(a)\leq m^{-1/2+\varepsilon/2},\max_a G_n(a)\leq n^{-1/2+\varepsilon/2}\right]$$
$$+\Pr_{H_0}\left\{\max_a G_m(a)>m^{-1/2+\varepsilon/2}\right\}+\Pr_{H_0}\left\{\max_a G_n(a)>n^{-1/2+\varepsilon/2}\right\},$$

where $G_m(a)=\sup_t\left|F_{Y(a)m}(t)-F_{H_0}(t;a)\right|,\ G_n(a)=\sup_t\left|F_{H_0}(t;a)-F_{X(a)n}(t)\right|$ provide

$$\sum_{k=1}^{4}G_{kimn}(a)\geq-2G_m(a)-2G_n(a)\ \text{and}\ \sum_{k=1}^{4}\Psi_{kjmn}(a)\geq-2G_m(a)-2G_n(a).$$

(The event $\left\{\max_a G_m(a)\leq m^{-1/2+\varepsilon/2},\max_a G_n(a)\leq n^{-1/2+\varepsilon/2}\right\}$ insures that the arguments of the log(.) 's appeared in the inequalities below are positive for large values of $n$.) This implies that, for

$$J_{1nm}(a)=\sum_{i=1}^{n}\log\left[\frac{n}{2n^{1-\gamma}(n+m)}\left\{\left(g_n\left(i+n^{1-\gamma}\right)-g_n\left(i-n^{1-\gamma}\right)\right)\left(1+mn^{-1}\right)-2mG_m(a)-2mG_n(a)\right\}\right],$$

$$J_{2nm}(a)=\sum_{j=1}^{m}\log\left[\frac{m}{2m^{1-\gamma}(n+m)}\left\{\left(g_n\left(j+m^{1-\gamma}\right)-g_n\left(j-m^{1-\gamma}\right)\right)\left(1+nm^{-1}\right)-2nG_m(a)-2nG_n(a)\right\}\right],$$

we have

$$\Pr_{H_0}\left\{(n+m)^{-1}\log\left(TS_{nm}\right)>C\right\}$$

$$\leq\Pr_{H_0}\left[(n+m)^{-1}\min_a\left\{J_{1nm}(a)+J_{2nm}(a)\right\}<-C,\ \max_a G_m(a)\leq m^{-1/2+\varepsilon/2},\max_a G_n(a)\leq n^{-1/2+\varepsilon/2}\right]$$
$$+\Pr_{H_0}\left\{\max_a G_m(a)>m^{-1/2+\varepsilon/2}\right\}+\Pr_{H_0}\left\{\max_a G_n(a)>n^{-1/2+\varepsilon/2}\right\}.$$

Now, we can evaluate probabilistic characteristics of notations containing $G_m(a)$ and $G_n(a)$ using Kolmogorov-Smirnov type inequalities. To this end, noting that

$g_n(t)=I(t\leq 0)+tI(0<t\leq n)+nI(t>n)$,

$$J_{1nm}(a)=\sum_{i=1}^{n^{1-\gamma}}\log\left[\frac{n}{2n^{1-\gamma}(n+m)}\left\{\left(i+n^{1-\gamma}-1\right)\left(1+mn^{-1}\right)-2mG_m(a)-2mG_n(a)\right\}\right]$$
$$+\sum_{i=n-n^{1-\gamma}+1}^{n}\log\left[\frac{n}{2n^{1-\gamma}(n+m)}\left\{\left(n-i+n^{1-\gamma}\right)\left(1+mn^{-1}\right)-2mG_m(a)-2mG_n(a)\right\}\right]$$
$$+\sum_{i=n^{\gamma}+1}^{n-n^{1-\gamma}}\log\left[\frac{n}{2n^{1-\gamma}(n+m)}\left\{2n^{1-\gamma}\left(1+mn^{-1}\right)-2mG_m(a)-2mG_n(a)\right\}\right]$$

and

$$J_{2nm}(a)=\sum_{j=1}^{m^{\gamma}}\log\left[\frac{m}{2m^{1-\gamma}(n+m)}\left\{\left(j+m^{1-\gamma}-1\right)\left(1+nm^{-1}\right)-2nG_m(a)-2nG_n(a)\right\}\right]$$
$$+\sum_{j=m-m^{\gamma}+1}^{m}\log\left[\frac{m}{2m^{1-\gamma}(n+m)}\left\{\left(n-j+m^{1-\gamma}\right)\left(1+nm^{-1}\right)-2nG_m(a)-2nG_n(a)\right\}\right]$$
$$+\sum_{j=m^{\gamma}+1}^{m-m^{\gamma}}\log\left[\frac{m}{2m^{1-\gamma}(n+m)}\left\{2m^{1-\gamma}\left(1+nm^{-1}\right)-2nG_m(a)-2nG_n(a)\right\}\right],$$

we obtain

$$\Pr_{H_0}\left\{(n+m)^{-1}\log(TS_{nm})>C\right\}$$
$$\le \Pr_{H_0}\left\{(n+m)^{-1}\left(\tilde J_{1nm}+\tilde J_{2nm}\right)<-C,\ \max_a G_m(a)\le m^{-1/2+\varepsilon/2}, \max_a G_n(a)\le n^{-1/2+\varepsilon/2}\right\}$$
$$+\Pr_{H_0}\left\{\max_a G_m(a)>m^{-1/2+\varepsilon/2}\right\}+\Pr_{H_0}\left\{\max_a G_n(a)>n^{-1/2+\varepsilon/2}\right\}$$

with

$$\tilde J_{1nm}=\sum_{i=1}^{n^{1-\gamma}}\log\left[\frac{i+n^{1-\gamma}-1}{2n^{1-\gamma}}-\frac{n^\gamma m}{n+m}\left\{\max_a G_m(a)+\max_a G_n(a)\right\}\right]$$
$$+\sum_{i=n-n^{1-\gamma}+1}^{n}\log\left[\frac{n-i+n^{1-\gamma}}{2n^{1-\gamma}}-\frac{n^\gamma m}{n+m}\left\{\max_a G_m(a)+\max_a G_n(a)\right\}\right]$$
$$+\sum_{i=n^\gamma+1}^{n-n^{1-\gamma}}\log\left[1-\frac{n^\gamma m}{n+m}\left\{\max_a G_m(a)+\max_a G_n(a)\right\}\right],$$

$$\tilde J_{2nm}=\sum_{j=1}^{m^\gamma}\log\left[\frac{j+m^{1-\gamma}-1}{2m^{1-\gamma}}-\frac{nm^\gamma}{n+m}\left\{\max_a G_m(a)+\max_a G_n(a)\right\}\right]$$
$$+\sum_{j=m-m^\gamma+1}^{m}\log\left[\frac{n-j+m^{1-\gamma}}{2m^{1-\gamma}}-\frac{nm^\gamma}{n+m}\left\{\max_a G_m(a)+\max_a G_n(a)\right\}\right]$$
$$+\sum_{j=m^\gamma+1}^{m-m^\gamma}\log\left[1-\frac{nm^\gamma}{n+m}\left\{\max_a G_m(a)+\max_a G_n(a)\right\}\right].$$

Thus, it is clear that, defining $0<\varepsilon<1-2\gamma$, where $\gamma<1/4$, we have, e.g., $(n+m)^{-1}n^\gamma m^{1/2+\varepsilon/2}\to 0$ as $n\to\infty$, where $m=O(n)$, and then, for relatively large $n$, we obtain

$$\Pr_{H_0}\left\{(n+m)^{-1}\log(TS_{nm})>C\right\}$$
$$\le I\left\{(n+m)^{-1}\left(C_{1nm}+C_{2nm}\right)<-C\right\}+\Pr_{H_0}\left\{\max_a G_m(a)>m^{-1/2+\varepsilon/2}\right\}\qquad \text{(A.2)}$$
$$+\Pr_{H_0}\left\{\max_a G_n(a)>n^{-1/2+\varepsilon/2}\right\},$$

where

$$C_{1nm}=\sum_{i=1}^{n^{1-\gamma}}\log\left(\frac{i+n^{1-\gamma}-1}{2n^{1-\gamma}}-\frac{n^\gamma m^{1/2+\varepsilon/2}}{n+m}-\frac{n^{\gamma-1/2+\varepsilon/2}m}{n+m}\right)$$
$$+\sum_{i=n-n^{1-\gamma}+1}^{n}\log\left(\frac{n-i+n^{1-\gamma}}{2n^{1-\gamma}}-\frac{n^\gamma m^{1/2+\varepsilon/2}}{n+m}-\frac{n^{\gamma-1/2+\varepsilon/2}m}{n+m}\right)+\left(n-2n^{1-\gamma}\right)\log\left(1-\frac{n^\gamma m^{1/2+\varepsilon/2}}{n+m}-\frac{n^{\gamma-1/2+\varepsilon/2}m}{n+m}\right)$$

$$\geq n^{1-\gamma}\log\left(\frac{1}{2}-\frac{n^{\gamma}m^{1/2+\varepsilon/2}}{n+m}-\frac{n^{\gamma-1/2+\varepsilon/2}m}{n+m}\right)+n^{1-\gamma}\log\left(\frac{1}{2}-\frac{n^{\gamma}m^{1/2+\varepsilon/2}}{n+m}-\frac{n^{\gamma-1/2+\varepsilon/2}m}{n+m}\right)$$
$$+\left(n-2n^{1-\gamma}\right)\log\left(1-\frac{n^{\gamma}m^{1/2+\varepsilon/2}}{n+m}-\frac{n^{\gamma-1/2+\varepsilon/2}m}{n+m}\right),$$

$$C_{2nm}=\sum_{j=1}^{m^{\gamma}}\log\left(\frac{j+m^{1-\gamma}-1}{2m^{1-\gamma}}-\frac{nm^{\gamma-1/2+\varepsilon/2}}{n+m}-\frac{n^{1/2+\varepsilon/2}m^{\gamma}}{n+m}\right)$$
$$+\sum_{j=m-m^{\gamma}+1}^{m}\log\left(\frac{n-j+m^{1-\gamma}}{2m^{1-\gamma}}-\frac{nm^{\gamma-1/2+\varepsilon/2}}{n+m}-\frac{n^{1/2+\varepsilon/2}m^{\gamma}}{n+m}\right)+\sum_{j=m^{\gamma}+1}^{m-m^{\gamma}}\log\left(1-\frac{nm^{\gamma-1/2+\varepsilon/2}}{n+m}-\frac{n^{1/2+\varepsilon/2}m^{\gamma}}{n+m}\right)$$

$$\geq m^{1-\gamma}\log\left(\frac{1}{2}-\frac{nm^{\gamma-1/2+\varepsilon/2}}{n+m}-\frac{n^{1/2+\varepsilon/2}m^{\gamma}}{n+m}\right)+m^{1-\gamma}\log\left(\frac{1}{2}-\frac{nm^{\gamma-1/2+\varepsilon/2}}{n+m}-\frac{n^{1/2+\varepsilon/2}m^{\gamma}}{n+m}\right)$$
$$+\left(m-2m^{\gamma}\right)\log\left(1-\frac{nm^{\gamma-1/2+\varepsilon/2}}{n+m}-\frac{n^{1/2+\varepsilon/2}m^{\gamma}}{n+m}\right)$$

and $\left(n+m\right)^{-1}\left(C_{1nm}+C_{2nm}\right)\rightarrow 0$ as $n\rightarrow\infty$.

Therefore, we conclude that

$$\Pr_{H_0}\left\{\left(n+m\right)^{-1}\log\left(TS_{nm}\right)>C\right\}\leq o(1)+\Pr_{H_0}\left\{\max_a G_m(a)>m^{-1/2+\varepsilon/2}\right\}$$
$$+\Pr_{H_0}\left\{\max_a G_n(a)>n^{-1/2+\varepsilon/2}\right\}.$$

Proposition 3 implies that we can define the set

$$D_{nm}=\left\{W_{ij}=\left(X_{1i}-Y_{1j}\right)\left(Y_{2j}-X_{2i}\right)^{-1},\ i=1,...,n,j=1,...,m\right\}$$
$$\cup\left\{U_{ij}=\left(X_{1i}-X_{1j}\right)\left(X_{2j}-X_{2i}\right)^{-1},\ 1\leq i\neq j\leq n\right\}$$
$$\cup\left\{V_{ij}=\left(Y_{1i}-Y_{1j}\right)\left(Y_{2j}-Y_{2i}\right)^{-1},\ 1\leq i\neq j\leq m\right\}$$

to present

$$\Pr_{H_0}\left\{\left(n+m\right)^{-1}\log\left(TS_{nm}\right)>C\right\}=\Pr_{H_0}\left\{\left(n+m\right)^{-1}\log\left(\max_a ELR_{X,n}(a)ELR_{Y,m}(a)\right)>C\right\}$$
$$=\Pr_{H_0}\left\{\left(n+m\right)^{-1}\log\left(\max_{a\in D_{nm}}ELR_{X,n}(a)ELR_{Y,m}(a)\right)>C\right\}.$$

Thus, $\Pr_{H_0}\left\{\left(n+m\right)^{-1}\log\left(TS_{nm}\right)>C\right\}\leq o(1)$

$$+\sum_{l=m,n}\left[\sum_{i=1}^{n}\sum_{j=1}^{m}\Pr_{H_0}\left\{G_l\left(W_{ij}\right)>l^{-1/2+\varepsilon/2},G_l\left(W_{ij}\right)=\max_{a\in D_{nm}}G_l\left(a\right)\right\}\right.$$

$$+\sum_{i=1}^{n}\sum_{j=1}^{n}\Pr_{H_0}\left\{G_l\left(U_{ij}\right)>l^{-1/2+\varepsilon/2},G_l\left(U_{ij}\right)=\max_{a\in D_{nm}}G_l\left(a\right)\right\}$$

$$\left.+\sum_{i=1}^{m}\sum_{j=1}^{m}\Pr_{H_0}\left\{G_l\left(V_{ij}\right)>l^{-1/2+\varepsilon/2},G_l\left(V_{ij}\right)=\max_{a\in D_{nm}}G_l\left(a\right)\right\}\right]$$

$$\le o(1)+\sum_{l=m,n}\left[\sum_{i=1}^{n}\sum_{j=1}^{m}\Pr_{H_0}\left\{G_l\left(W_{ij}\right)>l^{-1/2+\varepsilon/2}\right\}+\sum_{i=1}^{n}\sum_{j=1}^{n}\Pr_{H_0}\left\{G_l\left(U_{ij}\right)>l^{-1/2+\varepsilon/2}\right\}\right.$$

$$\left.+\sum_{i=1}^{m}\sum_{j=1}^{m}\Pr_{H_0}\left\{G_l\left(V_{ij}\right)>l^{-1/2+\varepsilon/2}\right\}\right]. \tag{A.3}$$

We consider

$$\theta=\sum_{l=m,n}\sum_{i=1}^{n}\sum_{j=1}^{m}\Pr_{H_0}\left\{G_l\left(W_{ij}\right)>l^{-1/2+\varepsilon/2}\right\}$$

that is a summand presented in the upper bound shown in (A.3). Note that, denoting $F^{-1}$ as the inverse or quantile function of a distribution function $F$, such that $F\left(F^{-1}\left(u\right)\right)=u$, we have

$$G_m(a)=\sup_u\left|F_{F_{H_0}^{-1}(Y(a);a)m}(u)-u\right|,\ \mathrm{G}_n(a)=\sup_u\left|u-F_{F_{H_0}^{-1}(X(a);a)n}\left(u\right)\right|$$

and then

$$\theta=\sum_{i=1}^{n}\sum_{j=1}^{m}\Pr_{H_0}\left[\sup_u\left|\frac{1}{m}I\left\{F_{H_0}^{-1}\left(Y_j(W_{ij});W_{ij}\right)\le u\right\}-\frac{1}{\left(m-1\right)m}\sum_{s=1,s\ne j}^{m}I\left\{F_{H_0}^{-1}\left(Y_s(W_{ij});W_{ij}\right)\le u\right\}\right.\right.$$
$$\left.\left.+\frac{1}{m-1}\sum_{s=1,s\ne j}^{m}I\left\{F_{H_0}^{-1}\left(Y_s(W_{ij});W_{ij}\right)\le u\right\}-u\right|>m^{-1/2+\varepsilon/2}\right]$$

$$+\sum_{i=1}^{n}\sum_{j=1}^{m}\Pr_{H_0}\left[\sup_u\left|\frac{1}{n}I\left\{F_{H_0}^{-1}\left(X_i(W_{ij});W_{ij}\right)\le u\right\}-\frac{1}{\left(n-1\right)n}\sum_{s=1,s\ne i}^{n}I\left\{F_{H_0}^{-1}\left(X_s(W_{ij});W_{ij}\right)\le u\right\}\right.\right.$$
$$\left.\left.+\frac{1}{n-1}\sum_{s=1,s\ne i}^{n}I\left\{F_{H_0}^{-1}\left(X_s(W_{ij});W_{ij}\right)\le u\right\}-u\right|>n^{-1/2+\varepsilon/2}\right],$$

where $W_{ij}=\left(X_{1i}-Y_{1j}\right)\left(Y_{2j}-X_{2i}\right)^{-1}$.

This leads to

$$\theta \le \sum_{i=1}^{n}\sum_{j=1}^{m} \Pr_{H_0}\left\{\tilde{G}_{1jm}\left(W_{ij}\right) > m^{-1/2+\varepsilon/2} - 2m^{-1}\right\} + \sum_{i=1}^{n}\sum_{j=1}^{m} \Pr_{H_0}\left\{\tilde{G}_{2in}\left(W_{ij}\right) > n^{-1/2+\varepsilon/2} - 2n^{-1}\right\},$$

where

$$\tilde{G}_{1jm}(a) = \sup_u \left| \frac{1}{m-1} \sum_{s=1, s\neq j}^{m} I\left\{F_{H_0}^{-1}\left(Y_s(a); a\right) \le u\right\} - u \right|,$$
$$\tilde{G}_{2in}(a) = \sup_u \left| \frac{1}{n-1} \sum_{s=1, s\neq i}^{n} I\left\{F_{H_0}^{-1}\left(X_s(a); a\right) \le u\right\} - u \right|.$$

It is clear that, when $0 < \varepsilon < 1-2\gamma$ and $\gamma < 1/4$, for relatively large $n$, such that $m^{-1/2+\varepsilon/2} - 2m^{-1} > bm^{-1/2+\varepsilon/2}$ and $n^{-1/2+\varepsilon/2} - 2n^{-1} > bn^{-1/2+\varepsilon/2}$ ($0 < b < 1$ is a constant), the inequality obtained above yields

$$\theta \le \sum_{i=1}^{n}\sum_{j=1}^{m} \mathrm{E}_{H_0} \Pr_{H_0}\left\{\tilde{G}_{1jm}\left(W_{ij}\right) > bm^{-1/2+\varepsilon/2} \mid W_{ij}\right\} + \sum_{i=1}^{n}\sum_{j=1}^{m} \mathrm{E}_{H_0} \Pr_{H_0}\left\{\tilde{G}_{2in}\left(W_{ij}\right) > bn^{-1/2+\varepsilon/2} \mid W_{ij}\right\},$$

where $\mathrm{E}_{H_0}$ means the expectation derived under $H_0$. For a fixed $W_{ij} = \left(X_{1i} - Y_{1j}\right)\left(Y_{2j} - X_{2i}\right)^{-1}$, the statistics $\tilde{G}_{1jm}\left(W_{ij}\right), \tilde{G}_{2in}\left(W_{ij}\right)$ contain the empirical distribution functions

$\left(m-1\right)^{-1} \sum_{s=1, s\neq j}^{m} I\left\{F_{H_0}^{-1}\left(Y_s(W_{ij}); W_{ij}\right) \le u\right\}$ and $\left(n-1\right)^{-1} \sum_{s=1, s\neq i}^{n} I\left\{F_{H_0}^{-1}\left(X_s(W_{ij}); W_{ij}\right) \le u\right\}$ that are based on independent and identically distributed random variables. By virtue of the theorem of Dvoretzky, Kiefer and Wolfowitz (Serfling, 2009, p. 59), we have, for $\varepsilon > 0$,

$$\Pr_{H_0}\left\{\tilde{G}_{1jm}\left(W_{ij}\right) > bm^{-1/2+\varepsilon/2} \mid W_{ij}\right\} \le C_1' \exp\left(-2b^2(m-1)m^{-1+\varepsilon}\right) \text{ and}$$
$$\Pr_{H_0}\left\{\tilde{G}_{2in}\left(W_{ij}\right) > bn^{-1/2+\varepsilon/2} \mid W_{ij}\right\} \le C_2' \exp\left(-2b^2(n-1)n^{-1+\varepsilon}\right),$$

where $C_1'$, $C_2'$ are finite positive constants (not depending on distributions of $X(a)$ and $Y(a)$).

Then $\theta \le C_1' \exp\left(-2b^2(m-1)m^{-1+\varepsilon}\right)nm + C_2' \exp\left(-2b^2(n-1)n^{-1+\varepsilon}\right)nm$.

It is clear that the terms presented in the upper bound obtained in (A.3) can be evaluated in a similar manner to the $\theta$'s evaluation shown above. This implies that, for constants $C_1 > 0, C_2 > 0$, we have

$$\begin{aligned}&\Pr_{H_0}\left\{(n+m)^{-1}\log\left(TS_{nm}\right) > C\right\}\\ &\qquad \le o(1) + C_1\exp\left(-2b^2(m-1)m^{-1+\varepsilon}\right)nm + C_2\exp\left(-2b^2(n-1)n^{-1+\varepsilon}\right)nm\end{aligned} \tag{A.4}$$

that means $\Pr_{H_0}\left\{(n+m)^{-1}\log\left(TS_{nm}\right) > C\right\} \to 0$ as $n \to \infty$ and Lemma A.2 is proven.

Lammas A.1 and A.2 provide the statement of Proposition 4.

## Proof of Proposition 5.

Without loss of generality, we can assume the framework that is presented above Lemma A.1 in the proof of Proposition 4. In this case, $X_i(a) = X_{1i} + aX_{2i}, i = 1,...,km,$ and $Y_i(a) = Y_{1i} + aY_{2i}, i = 1,...,km,$ where $a \in R^1$ and $k = 1,...,K$. Then,

$$R_{km} = \max_a ELR_{X,km}(a)ELR_{Y,km}(a).$$

Under $H_1$, by virtue of Lemma A.1, we can find a constant $C > 0$ such that, for $1 \le s \le K$, $\Pr_{H_1}\left\{\max_{1\le k\le K}\log\left(R_{km}\right) > mC\right\} \ge \Pr_{H_1}\left\{\log\left(R_{sm}\right) > mC\right\} \to 1$.

Under $H_0$, we have the inequality

$$\Pr_{H_0}\left\{\max_{1\le k\le K}\log\left(R_{km}\right) > mC\right\} \le \sum_{s=1}^{K}\Pr_{H_0}\left\{\log\left(R_{sm}\right) > mC\right\}.$$

Now, we reconsider the result (A.4), noting that the term $o(1)$ in Inequality (A.4) means $I\left\{(n+m)^{-1}\left(C_{1nm} + C_{2nm}\right) < -C\right\}$ used in (A.2), when the notations of the proof of Proposition 4 are in effect. Employing Result (A.4), where $n, m$ are redefined to equal to $sm$, we can show that, for $r = sm$,

$$\Pr_{H_0}\left\{m^{-1}\log\left(R_{sm}\right)>C\right\}$$
$$\le I\left\{m^{-1}\left(C_{1rr}+C_{2rr}\right)<-C\right\}+C_1\exp\left(-2b^2(sm-1)\left(sm\right)^{-1+\varepsilon}\right)s^2m^2$$
$$+C_2\exp\left(-2b^2(sm-1)\left(sn\right)^{-1+\varepsilon}\right)s^2m^2$$

with the deterministic term $m^{-1}\left(C_{1rr}+C_{2rr}\right)\to 0$, as $m\to\infty$, where $\varepsilon>0$, $0<b<1$ is a constant and $C_1$, $C_2$ are finite positive constants. Then, for large values of $sm$ and positive constants $M_1, M_2$, we can write

$$\Pr_{H_0}\left\{m^{-1}\log\left(R_{sm}\right)>C\right\}\le M_1 s^2 m^2\exp\left\{-M_2\left(sm\right)^{\varepsilon}\right\}.$$

This leads to the inequality

$$\Pr_{H_0}\left\{\max_{1\le k\le K}\log\left(R_{km}\right)>mC\right\}\le\sum_{s=1}^{K}\Pr_{H_0}\left\{\log\left(R_{sm}\right)>mC\right\}\le M_1K^3m^2\exp\left\{-M_2\left(m\right)^{\varepsilon}\right\}\to 0$$

that completes the proof of Proposition 5.

## R code to calculate the critical values of the proposed test (see Section 3.1)

```
library(MASS)

#Sample sizes

n1<-10  #n

n2<-10  #m

delta<-0.1

Test_Stat1<- array()

####### DBEL test statistic function ######################

DBEL_test1<- function(cc){

  txx1<- t(cc)%*%t(XX1)  #Vectors X and Y used in the paper in the code are  x1, x2

  txx2<- t(cc)%*%t(XX2)

  n<-n1+n2

  #rename data as x1 and x2 in the following analyses
```

```
x1<- txx1[1:n1]

x2<- txx2[1:n2]

sx<-sort(x1)

sy<-sort(x2)

###############################################

#######obtain the test statistic based on Sample 1############

###############################################

m<-c(round(n1^(delta+0.5)):min(c(round((n1)^(1-delta)),round(n1/2))))

a<-replicate(n1,m)

rm<-as.vector(t(a))

L<-c(1:n1)- rm

LL<-replace(L, L <= 0, 1 )

U<-c(1:n1)+ rm

UU<-replace(U, U > n1, n1)

xL<-sx[LL]

xU<-sx[UU]

F<-(UU-LL)+n2*(ecdf(sy)(xU)-ecdf(sy)(xL)) # the empirical distribution function

F<-F/(n1+n2)

F[F==0]<-1/(n1+n2)

I<-2*rm/ ( n1*F )

ux<-array(I, c(n1,length(m)))

tstat1<- log(min(apply(ux,2,prod)))

 ###################################################

#######obtain the test statistic based on Sample 2#################

###################################################

m<-c(round(n2^(delta+0.5)):min(c(round((n2)^(1-delta)),round(n2/2))))
```

```
  a<-replicate(n2,m)
  rm<-as.vector(t(a))
  L<-c(1:n2)- rm
  LL<-replace(L, L <= 0, 1 )
  U<-c(1:n2)+ rm
  UU<-replace(U, U > n2, n2)
  yL<-sy[LL]
  yU<-sy[UU]
  F<- n1*(ecdf(sx)(yU)-ecdf(sx)(yL)) + (UU-LL)
  F<-F/(n1+n2)
  F[F==0]<-1/(n1+n2)
  I<-2*rm/ ( n2*F )
  uy<-array(I, c(n2,length(m)))
  tstat2<- log(min(apply(uy,2, prod)))
  Test_Stat<- tstat1+tstat2
  return(Test_Stat)
}
#####################################################
MC<- 20000
for(mc in 1:MC){
  Sigma <- matrix(c(1,0,0,1),2,2)
  XX1<-mvrnorm(n = n1, rep(0, 2), Sigma)
  XX2<-mvrnorm(n = n2, rep(0, 2), Sigma)
  Ws<-c()
  XXX1<- XX1
  XXX2<- XX2
```

```
  XXX12<-rbind(XXX1,XXX2)
  for (i in 1:length(XXX12[,1])) {
    #since i != j, we delete the corresponding row
    dat_lz<- XXX12[-i,]
    for (j in 1:length(dat_lz[,1])) {
      lz1<- XXX12[i,]
      lz2<- dat_lz[j,]
      a2<- (lz1[1]-lz2[1])/(lz2[2]-lz1[2])
      Ws<- c(Ws,a2)
    }
  }
  Ws<- unique(Ws)
  Ws<- c(Ws,1,0)
  lz_a1<- c(rep(1,length(Ws)-2),0,1)
  xy_lab2<-cbind(lz_a1,Ws)
  stat<<-apply(xy_lab2,1,DBEL_test1)
  Test_Stat1[mc]<-max(stat)
  print(mc)
print(Test_Stat1[mc])
print(quantile(Test_Stat1,0.95))
}
```

## R code to calculate the critical values of the proposed test (see Section 4.1)

```
######################################################################
#### Group sequential critical values of two sample DBEL p=2###########
```

```
#####################################################################
library(MASS)
k<-   3  #total number of groups
m_k<- 5  #sample size per group
delta<-0.1
Test_Stat2<- array()
####### the DBEL test statistic functions ########
DBEL_test1<- function(cc){
  txx1<- t(cc)%*%t(XX1) #data used in the simulations
  txx2<- t(cc)%*%t(XX2)
  n<-n1+n2
  #rename the data as x1 and x2 in the following analyses
  x1<- txx1[1:n1]
  x2<- txx2[1:n2]
  sx<-sort(x1)
  sy<-sort(x2)
  #####################################################
  #######obtain the test statistic based on Sample 1################
  #####################################################
  m<-c(round(n1^(delta+0.5)):min(c(round((n1)^(1-delta)),round(n1/2))))
  a<-replicate(n1,m)
  rm<-as.vector(t(a))
  L<-c(1:n1)- rm
  LL<-replace(L, L <= 0, 1 )
  U<-c(1:n1)+ rm
  UU<-replace(U, U > n1, n1)
```

```
xL<-sx[LL]

xU<-sx[UU]

F<-(UU-LL)+n2*(ecdf(sy)(xU)-ecdf(sy)(xL))

F<-F/(n1+n2)

I<-2*rm/ ( n1*F )

ux<-array(I, c(n1,length(m)))

tstat1<-log(min(apply(ux,2,prod)))

###################################################

#######obtain the test statistic based on Sample 2#######

###################################################

m<-c(round(n2^(delta+0.5)):min(c(round((n2)^(1-delta)),round(n2/2))))

a<-replicate(n2,m)

rm<-as.vector(t(a))

#rm<-rep(m, each = n2)

L<-c(1:n2)- rm

LL<-replace(L, L <= 0, 1 )

U<-c(1:n2)+ rm

UU<-replace(U, U > n2, n2)

yL<-sy[LL]                ###obtain y(i-m)

yU<-sy[UU]                ###obtain y(i+m)

F<- n1*(ecdf(sx)(yU)-ecdf(sx)(yL)) + (UU-LL)

F<-F/(n1+n2)

F[F==0]<-1/(n1+n2)

I<-2*rm/ ( n2*F )

uy<-array(I, c(n2,length(m)))

tstat2<-log(min(apply(uy,2, prod)))
```

```
  #log of the density-based EL ratio test statistic
  Test_Stat<- tstat1+tstat2
  return(Test_Stat)
}
MC<-20000 # number of iterations; set 1 for testing purpose
for(mc in 1:MC){
  #Total sample sizes
  n1<-k*m_k
  n2<-k*m_k
  Sigma <- matrix(c(1,0,0,1),2,2)
  ## Data simulated are stacked in rows.
  X<-mvrnorm(n = n1, rep(0, 2), Sigma)
  Y<-mvrnorm(n = n2, rep(0, 2), Sigma)

  group_stat<-array()
  for (mc2 in 1:k) {
    n1<- mc2*m_k #the sequential adaptive sample size
    n2<- mc2*m_k
    XX1<-X[1:n1,]
    XX2<-Y[1:n2,]
    #2. Compute W's the coefficients
    ## Compute W's, the coefficients u_2's.
    Ws<-c()
    ## New data to compute Ws and to keep raw data intact
    XXX1<- XX1
    XXX2<- XX2
```

```
    XXX12<-rbind(XXX1,XXX2)
    for (i in 1:length(XXX12[,1])) {
      #since i != j, we delete the corresponding row
      dat_lz<- XXX12[-i,]
      for (j in 1:length(dat_lz[,1])) {
        lz1<- XXX12[i,]
        lz2<- dat_lz[j,]
        a2<- (lz1[1]-lz2[1])/(lz2[2]-lz1[2])
        Ws<- c(Ws,a2)
      }
    }
    # the end step for W's is to add the case (u_1=0,u_2=1)
    Ws<- unique(Ws)
    Ws<- c(Ws,1,0)
    lz_a1<- c(rep(1,length(Ws)-2),0,1)
    xy_lab2<-cbind(lz_a1,Ws)
    #3. Compute the values of test statistics
    stat<<-apply(xy_lab2,1,DBEL_test1)
    group_stat[mc2]<-max(stat)
  }
  print(group_stat)
  Test_Stat2[mc]<- max(group_stat)
  print(mc)
print(Test_Stat2[mc])
print(quantile(Test_Stat2,0.95))
}
```